\documentclass[twocolumn,aps,pra,longbibliography,superscriptaddress,nofootinbib,10pt]{revtex4-1}
\usepackage[latin1]{inputenc}
\usepackage{graphicx,dcolumn,bm}
\usepackage{subfigure}
\usepackage{multirow}
\usepackage{array}
\usepackage{arydshln}
\usepackage{color}
\usepackage[colorlinks,linkcolor=blue,citecolor=blue,hyperindex]{hyperref}
\usepackage{amsfonts}
\usepackage{amssymb}
\usepackage{amsmath}
\usepackage{latexsym}
\usepackage{simplewick}
\usepackage{times,txfonts}
\usepackage{epstopdf}
\usepackage{multirow}
\usepackage{float}
\usepackage[table]{xcolor}
\usepackage{multibib}
\usepackage{mathrsfs}
\usepackage[sans]{dsfont}
\usepackage[mathscr]{eucal}
\usepackage{soul}
\usepackage{epsfig}
\definecolor{Blue}{rgb}{0.0,0.0,1}
\definecolor{Red}{rgb}{1,0.0,0.0}
\definecolor{Green}{rgb}{0,0.5,0.0}
\newcommand{\ket}[1]{{|{#1}\rangle}}

\newcommand{\vac}{\tilde 0}

\usepackage{tikz}
\usetikzlibrary{decorations.pathmorphing}
\usetikzlibrary{arrows}
\usetikzlibrary{intersections,shapes.arrows}
\usetikzlibrary{calc}
\usetikzlibrary{quotes,angles}
\usepackage{nicefrac}
\usepackage{pgfplots}
\usepgfplotslibrary{fillbetween}
\pgfplotsset{compat=1.13,colormap={violetnew}{rgb=(0.293416, 0.0574044, 0.529412) rgb=(0.394818,0.233715,0.671945) rgb =(0.49622,0.410025,0.814477) rgb=(0.588672,0.567494,0.910066) rgb=(0.663226,0.687282,0.911765) rgb=(0.73778,0.807069,0.913465) rgb=(0.807267,0.861883,0.894034) rgb=(0.874222,0.884211,0.864039) rgb=(0.941176, 0.906538, 0.834043)}}
\usepgfplotslibrary{groupplots} 
\usepgfplotslibrary[groupplots] 
\usetikzlibrary{pgfplots.groupplots} 
\usetikzlibrary[pgfplots.groupplots] 
\usepgfplotslibrary{statistics}
\usepackage{pgfplotstable}
\tikzset{jumpdot/.style={mark=*,solid},excl/.append style={jumpdot,fill=white},incl/.append style={jumpdot,fill=black}}

\begin{document}

\title{Dynamics of quantum resources in regular and Majorana fermion systems}

\author{Diego Paiva Pires}
\affiliation{Departamento de F\'{i}sica, Universidade Federal do Maranh\~{a}o, Campus Universit\'{a}rio do Bacanga, 65080-805, S\~{a}o Lu\'{i}s, Maranh\~{a}o, Brazil}
\author{Diogo O. Soares-Pinto}
\affiliation{Instituto de F\'{i}sica de S\~{a}o  Carlos, Universidade de S\~{a}o Paulo, CP 369, 13560-970, S\~{a}o Carlos, SP, Brazil}
\author{E. Vernek}
\affiliation{Instituto de F\'{i}sica, Universidade Federal  de Uberl\^{a}ndia, 38400-902 Uberl\^{a}ndia, Minas Gerais, Brazil}
\begin{abstract}
In recent years, the reassessment of quantum physical phenomena under the framework of resource theories has triggered the design of novel quantum technologies that take advantage from quantum resources, such as entanglement and quantum coherence. Bearing this in mind, in this work we study the dynamics of quantum resources for two solid-state fermionic quantum devices: (i) a system composed by a pair of Majorana fermions and (ii) another comprising a pair of regular fermions. In both systems, the fermionic species are coupled to a single-level quantum dot. From the interaction of these tripartite systems with a dissipative reservoir, we were able to characterize the dynamics of the devices for some initial states. By employing a time-nonlocal master equation approach, we obtain the evolution for fermionic occupations, quantum correlations, and quantum coherences in both the Markovian and non-Markovian dissipating regimes. We investigate the interconversion of local coherence and bipartite correlations for the marginal states in each device. While the dynamics of the entanglement and quantum coherence depend quite strongly on the temperature of the reservoir for regular fermions, we found these evolutions are qualitatively similar for the case of Majorana bound states regardless of temperature. Our results illustrate the use of quantum information-theoretic measures to characterize the role of quantum resources in fermion systems.
\end{abstract}

\maketitle


\section{Introduction}
\label{sec:sec0001}

Quantum information science has paved the way for a comprehensive understanding of quantum phenomena as resources, which in turn may be consumed to perform tasks that are typically not possible otherwise~\cite{RevModPhys.91.025001}. This is the case of entanglement and quantum coherence, both concepts being rigorously formulated under the framework of resource theories~\cite{PhysRevLett.122.120503,PhysRevResearch.2.012035,PhysRevLett.119.140402,PhysRevLett.116.120404,YungerHalpern2017}. Indeed, the former is an essential resource to enhance precision of phase estimation tasks in quantum me\-tro\-lo\-gy~\cite{PhysRevLett.102.100401}, also useful for quantum key distribution~\cite{PhysRevLett.67.661,PhysRevLett.68.557}, while the later finds applications in quantum optics~\cite{RevModPhys.37.231}, quantum thermodynamics~\cite{PhysRevA.93.052335}, and many-body physics~\cite{PhysRevB.93.184428}. Due to such characteristics, electronic and spin states have emerged as very promising platforms to design high-sensitive devices with applicability from materials science to biochemistry~\cite{RevModPhys.89.035002,PhysRevLett.126.170404,DeMille990,natrevmats.2017.88,Aslam67,nature25781,annurev-physchem-040513-103659}. Hence, designing physical systems that take advantage from quantum resources in a controllable fashion becomes imperative.

The ubiquitous idea proposed by Majorana~\cite{Majorana_1937} regarding particles that constitutes their own antiparticles, nowadays called Majorana fermions, opened new avenues in several areas of physics, including quantum computation~\cite{KITAEV20032,PhysRevB.97.205404,PhysRevX.6.031016,RevModPhys.80.1083,PhysRevLett.117.120403,PhysRevA.101.032106,QIP_0981_2014}. Unlike ordinary Dirac fermions such as electrons and protons, i.e., spin-${1}/{2}$ charged particles described by complex fields, Majorana fermions are spin-${1}/{2}$ neutral particles whose real field equations remains invariant to charge conjugation symmetry~\cite{NaturePhys.5.614.2009}. Over the past 80 years, pro\-bing the signature of Majorana fermions still remains an experimental challenge for the high-energy physics community. Quite recently, Majorana fermions has been predicted as zero-mode quasiparticle excitations in quantum many-body systems and its signatures experimentally observed in some solid-state setups~\cite{PhysRevLett.104.040502,0034-4885/75/7/076501,PhysRevB.81.125318,PhysRevB.61.10267,PhysRevB.91.081405,PhysRevLett.105.177002,10.1126/science.1222360,10.1021/nl303758w,10.1126/science.1259327,10.1126/science.aaf3961,10.1038/nphys2479,PhysRevB.98.085125,PhysRevB.104.115415}. Indeed, Majorana fermions play an important role as quasiparticle excitation in prototypical models for fault-tolerant topological quantum computation~\cite{KITAEV20032}, possibly manipulated in topological superconductors~\cite{PhysRevB.61.10267}, or fractional quantum Hall systems~\cite{MOORE1991362}, thermally induced topological phase transitions~\cite{PhysRevB.86.155140,PhysRevB.88.155141,PhysRevLett.112.130401,PhysRevLett.113.076408,njpQI_4_10_2018_0056,2106.05988}, and even in driven dissipative devices~\cite{PhysRevB.102.134501,PhysRevLett.125.147701}. Noteworthy, the interest in Majorana fermions relies on their exotic properties, such as non-Abelian statistics, thus differing quite radically from the original conventional electrons that condense into the superconducting state~\cite{PhysRevLett.98.237002}. Despite the intense debate regarding some experimental results in recent years, the hope for rea\-li\-zing quantum devices based on  Majorana fermions is still alive~\cite{d41586-021-00612-z,d41586-021-00954-8}.

Motivated by the potential applicability of Majorana fermions (MFs), in this work we study the dynamics of two fermionic systems, one composed of MFs and the other comprising regular fermions (RFs). Both systems are mediated by a quantum dot (QD) in an experimentally feasible solid-state setup~\cite{PhysRevB.104.115415,SciRep_You_2014}. From an open quantum dynamics approach, we investigate the behavior of the quantum resources of these fermionic systems combined to the QD, i.e., quantum coherence and entanglement, thus showing these quantities are somehow correlated. By consi\-de\-ring initial states with different fermionic occupations, the results show the two-body marginal states of each system exhibit nonzero values of concurrence and quantum coherences du\-ring the nonunitary dynamics. Moreover, the time evolution of such quantum resources behaves quite distinctly when the QD is coupled to MFs as compared to the case in which it is coupled to RFs. While in the former configuration the quantum resources evolve quite similarly regardless of temperature, in the latter they are rather distinct.

We find that quantum coherence and entanglement coincide for the case of the dynamics of regular fermions at zero temperature, for most of the initial states that have been considered. Our results indicate that quantum coherence can be consumed and converted into entanglement and vice versa. This interconversion process can occur for both the systems of Majorana fermions and regular fermions, at zero and finite temperatures. In detail, the local coherence (correlations) is consumed and partially transformed into local bipartite correlations (quantum coherence) in the system formed by MFs$+$QD, RFs$+$QD, and MFs$+$RFs.

\begin{figure}[t]
\begin{center}
\includegraphics[scale=0.8]{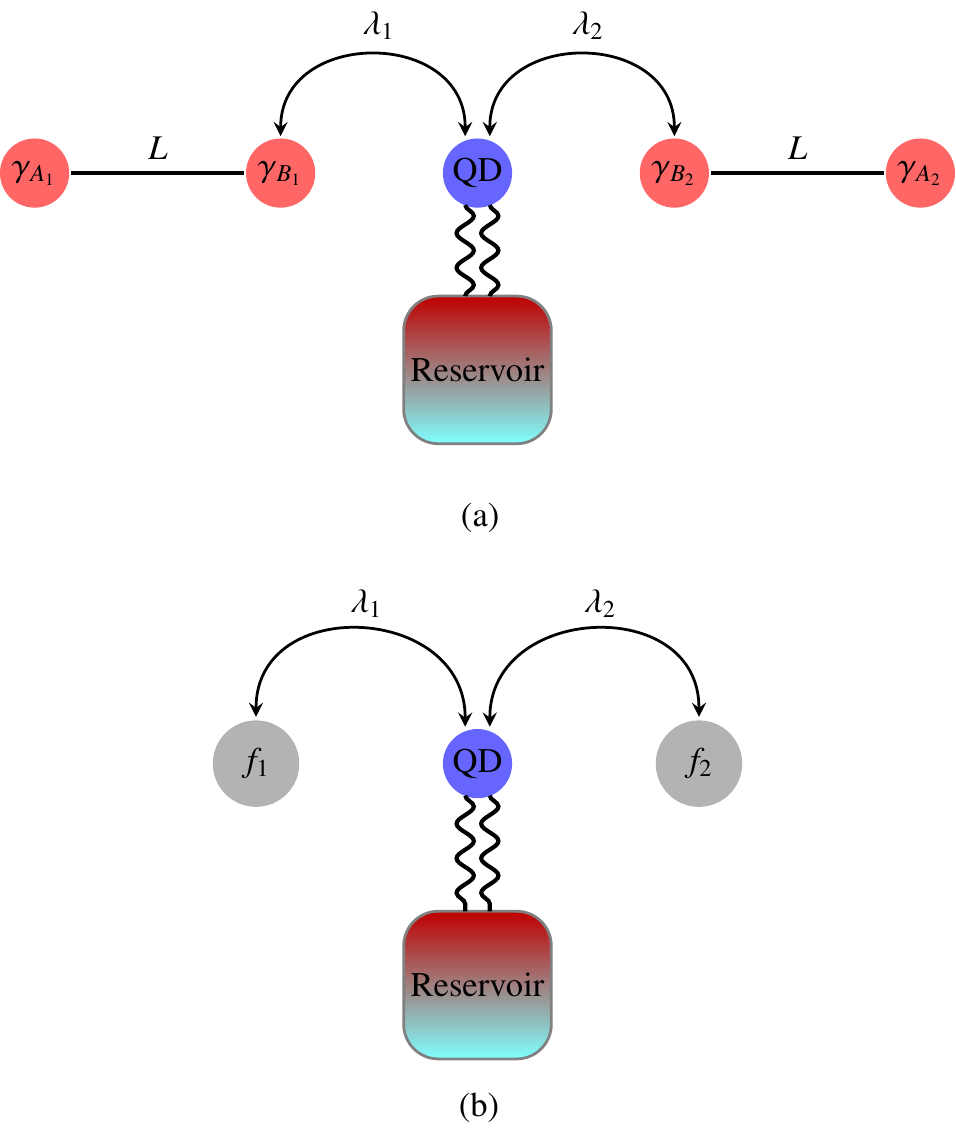}
\caption{(Color online) Schematic representation of the physical setups discussed in the paper. In the upper panel (a), the nonlocal Majorana fermions appear as edge modes in the ends of two nanowires of length $L$. The pair of MFs ${\gamma_{B_{1}}}$ and ${\gamma_{B_{2}}}$ are coupled to a single-level quantum dot (QD), with $\lambda_1$ and $\lambda_2$ being the coupling parameters of the QD to its $j$th neighboring MFs. The QD interacts with a fermionic reservoir at finite temperature. The lower panel (b) depicts the system of two electronic $f$ orbitals, i.e., regular fermions, which in turn hybridizes with the electronic orbitals of the quantum dot by means of coupling constants $\lambda_1$ and $\lambda_2$. The system of RFs$+$QD is allowed to interact to a bath of free fermions, thus modeling an open quantum system.}
\label{fig:fig00001}
\end{center}
\end{figure}

The paper is organized as follows. In Sec.~\ref{sec:sec0002}, we introduce the details of the physical models. In Sec.~\ref{sec:sec0003}, we describe the open system dynamics due to the coupling of the physical models with a dissipative environment. Such environment depends on the ohmicity parameter in a way that we can set both the Markovian and non-Markovian regimes. In Sec.~\ref{sec:sec0004n}, we analyze the dynamics of en\-tan\-glement and quan\-tum co\-he\-rences in the system of MFs$+$QD and RFs$+$QD under the influence of the environment in both regimes, at zero and finite temperatures. In Sec.~\ref{sec:sec0004a}, we present numerical analysis of occupations for a separable initial state with single-fermion occupation. In Sec~\ref{sec:sec0004b}, we discuss the dynamics of concurrence and $\ell_1$ norm of coherence for entangled initial states with single-fermion and two-fermion occupations. Focusing on the system of MFs$+$QD, we discuss in Sec.~\ref{sec:sec0005} the role of nonlocality of the Majorana bound states in the dynamics of occupations and quantum resources. Finally, in Sec.~\ref{sec:conclusions} we summarize our conclusions.


\section{The Model}
\label{sec:sec0002}
In this section we will investigate the spectral properties of the two different models depicted in Fig.~\ref{fig:fig00001}: the first one consists of nonlocal Majorana bound states, while the second comprises canonical regular fermions. In both systems, the two fermionic species couples to a single level quantum dot (QD). Throughout this work we assume a system of spinless fermions, which can be experimentally realized by applying a strong magnetic field. In fact, spinless fermion is an important condition for the emergence of Majorana bound states in topological superconductors~\cite{MOORE1991362,PhysRevB.81.125318,0034-4885/75/7/076501}. Unless other\-wise stated, the coupling to the reservoir is turned off and thus both MFs$+$QD and RFs$+$QD setups can be understood as closed quantum systems. The open quantum system scenario will be investigated in Sec.~\ref{sec:sec0003}.


\subsection{Majorana fermions}
\label{subsec:sec0002a}
%
For the case of Majorana fermions, we consider two wires with finite length $L$, supporting Majorana modes in their ends. Figure~\ref{fig:fig00001}(a) depicts this physical setting. In our system, the Majorana modes emerge at the edges of the pair of nanowires disposed in a serial geometric arrangement. We assume that the superconducting wires are far apart from each other, both of them in the topological regime. This proposed structure is similar to the setup experimentally realized using InAs/InP heterostructures in Ref.~\cite{PhysRevB.104.115415}. Here we assume that both the source and drain wire segments are co\-ve\-red by supercondutors, allowing for topological superconductivity in both sides of the system. Because for finite length of the wires, we shall expect a nonzero overlap between the two MF wave functions at the ends of each nanowire. Within this geometry, we neglect direct overlap of leftmost and rightmost wave functions. The Hamiltonian of the paired Majorana fermions is given by
\begin{equation}
 \label{eq:eq000001}
{H_{MF}} = \frac{i}{2}\, {\sum_{j = 1,2}} \, {\epsilon_{j}}\, {\gamma_{A_j}}{\gamma_{B_j}} ~,
\end{equation}
where $\gamma_{{A_j},{B_j}}$ are the two MF operators satisfying the Clifford algebra $\{ {\gamma_{X_j}}, {\gamma_{Y_l}}\} = 2 \,{\delta_{X,Y}}\,{\delta_{j,l}}$ and ${\epsilon_j}  \sim e^{-L/{\xi_j}}$ stands for the hybridization between each pair of MF edge modes due to the overlap of their wavefunctions, with $\xi_j$ being the effective coherence length~\cite{Wang_2013}. Importantly, the coupling energy $\epsilon_j$ decreases exponentially with the length $L$ of the nanowires, thus vanishing in the case of $L \gg {\xi_j}$. Furthermore, the MF pair becomes equivalent to a single zero-energy regular fermion (RF) in the asymptotic regime $L \rightarrow \infty$. Next, we will consider the single level non-interacting quantum dot (QD) modeled by the Hamiltonian
\begin{equation}
 \label{eq:eq000002}
{H_{QD}} = {\epsilon_d}\, {\hat{n}_d} ~,
\end{equation}
where $\epsilon_d$ is the energy of the QD and ${\hat{n}_d} = {d^{\dagger}}d$ is the electron number operator, while $d$ ($d^{\dagger}$) is the electron annihilation (creation) operator. The MF modes ${\gamma_{B_{1}}}$ and ${\gamma_{B_{2}}}$ in each wire leak into the QD and modify its electronic properties~\cite{PhysRevB.89.165314}. The coupling between the QD and the pair of Majorana edge modes ${\gamma_{B_{1}}}$ and ${\gamma_{B_{2}}}$ is described by the tunneling Hamiltonian
\begin{equation}
 \label{eq:eq000003}
{H_{MD}} = {\sum_{j = 1,2}}\, {\lambda_j}({d^{\dagger}} - d)\, {\gamma_{Bj}} ~,
\end{equation}
where the real parameters $\{ \lambda_j \}_{j = 1,2}$ characterize the coupling of the QD to its $j$th neighboring MFs.

Remarkably, MF operators can be mapped onto RF creation and annihilation operators as ${\gamma_{A_j}} = i({f_j} - {f_j^{\dagger}})$ and ${\gamma_{B_j}} = {f_j} + {f_j^{\dagger}}$, where $f_j$ ($f_j^{\dagger}$) is the RF annihilation (creation) operator fulfilling the anticommutation relations $\{{f_j}, {f_l^{\dagger}}\} = {\delta_{jl}}$, and $\{{f_j},{f_l}\} = \{{f_j^{\dagger}}, {f_l^{\dagger}}\}  = 0$. Noteworthy, RFs are nonlocal in the sense that they are composed by Majorana modes living far apart from each other. Next, it can be readily shown that the Hamiltonian of paired MFs in Eq.~\eqref{eq:eq000001} can be recast as
\begin{equation}
\label{eq:eq000004}
{H_{MF}} = {\sum_{j = 1,2}} \, {\epsilon_j}\left({\hat{n}_j} - \frac{1}{2}\right) ~,
\end{equation}
where ${\hat{n}_j} = {f^{\dagger}_j}{f_j}$ is the RF number operator. Similarly, the Hamiltonian in Eq.~\eqref{eq:eq000003} which models the QD-MF coupling is rewritten as
\begin{equation}
\label{eq:eq000005}
{H_{MD}} = {\sum_{j = 1,2}}{\lambda_j}({d^{\dagger}}{f_j} + {f_j^{\dagger}}d  + {d^{\dagger}}{f_j^{\dagger}} + {f_j}d) ~.
\end{equation}
Finally, combining Eqs.~\eqref{eq:eq000002},~\eqref{eq:eq000004}, and~\eqref{eq:eq000005}, the total Hamiltonian of the system reads
\begin{equation}
\label{eq:eq000006}
{H_M} = {\epsilon_d}{\hat{n}_d} + {\sum_{j = 1,2}}{\epsilon_j}\left({\hat{n}_j} - \frac{1}{2}\right) + {\sum_{j = 1,2}}{\lambda_j}({d^{\dagger}}{f_j} + {d^{\dagger}}{f_j^{\dagger}} + \text{H.c.}) ~.
\end{equation}


\subsection{Regular fermions}
\label{subsec:sec0002b}

For the system composed of regular fermions, we assume the QD is coupled to two other electron orbitals as depicted in Fig.~\ref{fig:fig00001}(b). These orbitals could be thought of as two other quantum dots. For the sake of clarity, we will continue using the same notation of $\{ {f_j} \}_{j = 1,2}$ operators for these orbitals. In this case, the Hamiltonian of the system is written as 
\begin{equation}
\label{eq:eq000007}
{H_R} = {\epsilon_d}{\hat{n}_d} + {\sum_{j = 1,2}}{\epsilon_j}\left({\hat{n}_j} - \frac{1}{2}\right) + {\sum_{j = 1,2}}{\lambda_j}({d^{\dagger}}{f_j} + {f_j^{\dagger}}{d}) ~,
\end{equation}
where the first term describes the QD, the second describes the electron orbitals, and the third accounts for the hybridization between the QD and the $f$ orbitals. Similar to the previous case, here we have ${\hat{n}_d} = {d^{\dagger}}d$ and ${\hat{n}_j} = {f^{\dagger}_j}{f_j}$ for $j = \{1,2\}$. Note that, for convenience, the energy of the $f$ orbitals is shifted by the constant value $(-1/2){\sum_{j=1,2}}{\epsilon_j}$, which does not change the dynamics of the quantum states.


\subsection{Generalized Hamiltonian}
\label{subsec:sec0002c}

Next, we will present a generalized Hamiltonian that encompasses both MF and RF systems. From Eqs.~\eqref{eq:eq000006} and~\eqref{eq:eq000007} we see that, apart from the terms ${d^{\dagger}} {f_j^{\dagger}} + {f_j}\, d$ inherited from the tunnel coupling of the QD and MFs, both Hamiltonians have the same form. It is therefore convenient to define a generic Hamiltonian that comprises both MF and RF systems as
\begin{equation}
\label{eq:eq000008}
{H_{M,R}} = {\epsilon_d}{\hat{n}_d} + {\sum_{j = 1}^2}\, {\epsilon_j}\left({\hat{n}_j} -  \frac{1}{2}\right) + {\sum_{j = 1,2}}({\lambda_j}\,{d^{\dagger}}{f_j} + {\widetilde{\lambda}_j}\,{d^{\dagger}}{f_j^{\dagger}} + {\rm H.c.}) ~.
\end{equation}
In particular, the Hamiltonian describing the MF system [see Eq.~\eqref{eq:eq000006}] is recovered by choosing ${\widetilde{\lambda}_j} = {\lambda_j}$, while one modeling the RF system [see Eq.~\eqref{eq:eq000007}] is obtained by setting ${\widetilde{\lambda}_j} = 0$.

In the following we discuss the matrix representation of the generalized Hamiltonian with respect to the basis $\{ |{n_1},{n_2}, {n_d}\rangle\}$, where the index ${n_{1,2,d}} = \{ 0,1 \}$ assigns the occupation number in the single-particle states created with the operators $f^{\dagger}_1$, $f^{\dagger}_2$, and $d^{\dagger}$ acting on the vacuum state $\ket{\tilde 0}\equiv |{0_1},{0_2},{0_d}\rangle$. To study in detail the Hamiltonian in Eq.~\eqref{eq:eq000008}, we fix the basis ordering $\{ \ket{\vac}$, ${f_1^{\dagger}} {d^{\dagger}}\ket{\vac}$, ${f_2^{\dagger}} {d^{\dagger}}\ket{\vac}$, ${f_1^{\dagger}}{f_2^{\dagger}}\ket{\vac}, {d^{\dagger}}\ket{\vac}$, ${f_1^{\dagger}}\ket{\vac}$, ${f_2^{\dagger}}\ket{\vac}$, ${f_1^{\dagger}} {f_2^{\dagger}} {d^\dagger} \ket{\tilde 0} \}$. Note that the first four states are eigenvectors of the operator $\hat N = {\hat{n}_1} + {\hat{n}_2} + {\hat{n}_d}$ with even eigenvalues, while for the last four states the eigenvalues are odd. We see that all processes described by the generic Hamiltonian either conserve the number of particles (for the RF system) or change it by two particles (for the MF system). With respect to this basis, the Hamiltonian in Eq.~\eqref{eq:eq000008} takes the form
\begin{equation}
\label{eq:eq000010}
H = \frac{1}{2}\left(\mathbb{I} + {\sigma_z} \right) \otimes {\mathcal{H}_0} + \frac{1}{2}\left(\mathbb{I} - {\sigma_z}\right)\otimes {\mathcal{H}_1} ~,
\end{equation}
where $\mathbb{I}$ is the $2\times 2$ identity matrix and  $\sigma_z$ is the Pauli matrix, with the Hermitian blocks
\begin{equation}
\label{eq:eq000011}
{\mathcal{H}_0} = \left[\begin{matrix} -{\epsilon_+} & -{\widetilde\lambda_1} & -\widetilde\lambda_2 & 0\\ 
- {\widetilde\lambda_1} & {\epsilon_d} + {\epsilon_-} & 0 & {\lambda_2} \\ 
-{\widetilde\lambda_2} & 0 & {\epsilon_d} - {\epsilon_-} & -{\lambda_1} \\ 
0 & {\lambda_2} & -{\lambda_1} & {\epsilon_+} \end{matrix}\right] ~,
\end{equation}
and
\begin{equation}
\label{eq:eq000012}
{\mathcal{H}_1} = \left[\begin{matrix} {\epsilon_d} - {\epsilon_+} & {\lambda_1} & {\lambda_2} & 0\\ 
{\lambda_1} & {\epsilon_-} & 0 & - {\widetilde{\lambda}_2} \\ 
{\lambda_2} & 0 & - {\epsilon_-} & {\widetilde{\lambda}_1} \\ 
0 & - {\widetilde{\lambda}_2} & {\widetilde{\lambda}_1} & {\epsilon_d} + {\epsilon_+} 
\end{matrix}\right] ~,
\end{equation}
where we have defined ${\epsilon_{\pm}}  = ({\epsilon_1} \pm {\epsilon_2})/2$.
\begin{table}[!t]
\caption{Eigenvectors and eigenvalues of the Hamiltonian in Eq.~\eqref{eq:eq000010} for the case of Majorana fermions (${\widetilde{\lambda}_j} = {\lambda_j}$). Here we have defined the parameters $\xi_{\pm}=(\epsilon_d\pm\epsilon)/2$ and ${\Delta_{\pm}} = \sqrt{{\xi_{\pm}^2} + {\lambda^2_1}+\lambda_2^2}$, while ${b_{\mu\nu}^{-1}} = {\sqrt{\,2{\Delta_{\mu}}({\Delta_{\mu}} + \nu{\xi_{\mu}})}}$ and ${c_{\mu\nu}} = ({\xi_{\mu}} + \nu{\Delta_{\mu}}){b_{\mu\nu}}$, with $\mu, \nu=\{+,- \}$.}
\begin{center}
\begin{tabular}{ll}
\hline\hline
Eigenstate & Energy  \\
\hline
$|{E_1}\rangle = \left({b_{++}}({\lambda_2}{f_2^{\dag}} + {\lambda_1}{f_1^\dag}){d^\dag} + {c_{++}} \right)\ket{\vac}$ & ${E_1} = {\xi_-} - {\Delta_+}$ \\
$|{E_2}\rangle = \left({b_{--}}({\lambda_1} {f_1^\dag} + {\lambda_2}{f_2^\dag}) + {c_{--}}{d^\dag}\right)\ket{\vac}$ & ${E_2} = {\xi_-} - {\Delta_-}$  \\
$|{E_3}\rangle = \left({b_{+-}} ({\lambda_1}{f^\dag_2} - {\lambda_2}{f^\dag_1}) + {c_{+-}}{f_1^\dag}{f_2^\dag}{d^\dag}\right)\ket{\vac} $ & ${E_3} = {\xi_+} - {\Delta_+}$  \\
$|{E_4}\rangle = \left({b_{-+}}({\lambda_1}{f^\dag_2} - {\lambda_2}{f^\dag_1}){d^\dag} + {c_{-+}}{f^{\dag}_1}{f^{\dag}_2}\right)\ket{\vac}$ & ${E_4} = {\xi_+} - {\Delta_-}$  \\
$|{E_5}\rangle = \left({b_{-+}}({\lambda_2}{f^\dag_2} + {\lambda_1}{f^\dag_1}) + {c_{-+}}{d^\dag}\right)\ket{\vac}$ & ${E_5} = {\xi_-} + {\Delta_-}$ \\
$|{E_6}\rangle = \left({b_{+-}}({\lambda_1}{f^\dag_1} + {\lambda_2}{f^\dag_2}){d^\dag} + {c_{+-}}\right)\ket{\vac}$ & ${E_6} = {\xi_-} +{\Delta_+}$ \\
$|{E_7}\rangle = \left({b_{--}}({\lambda_2} {f^\dag_1} - {\lambda_1}{f^\dag_2}){d^\dag} - {c_{--}}{f^\dag_1}{f^\dag_2}\right)\ket{\vac}$ & ${E_7} = {\xi_+} + {\Delta_-}$ \\
$|{E_8}\rangle = \left({b_{++}}({\lambda_1}{f^\dag_2} - {\lambda_2}{f^\dag_1}) + {c_{++}}{f^\dag_1}{f^\dag_2}{d^\dag}\right)\ket{\vac}$ & ${E_8} = {\xi_+} + {\Delta_+}$ \\
\hline\hline
\end{tabular}
\label{tab:tab000001}
\end{center}
\end{table}

Next, we will discuss the spectral properties of the Hamiltonian in Eq.~\eqref{eq:eq000010} for the case of MFs and RFs. For simplicity, from now on we will focus on the particular case of ${\epsilon_1} = {\epsilon_2} = \epsilon$, which implies ${\epsilon_-} = 0$ and ${\epsilon_+} = \epsilon$. For the system of MFs, we set ${\widetilde{{\lambda}_j}} = {{\lambda}_j}$ in Eqs.~\eqref{eq:eq000011} and~\eqref{eq:eq000012}. The set of eigenstates ${\{|{E_j}\rangle\}_{j = 1,\ldots,8}}$ and energies ${\{ {E_j} \}_{j = 1,\ldots,8}}$ of the Hamiltonian for MFs are listed in Table~\ref{tab:tab000001}. It is worth of mentioning that ${E_1}$ corresponds to the ground state of the system. Importantly, the set of states $\{|{E_j}\rangle \}_{j = 1,\ldots,4}$ exhibits even parity respective to the occupation number, while the set $\{|{E_j}\rangle \}_{j = 5,\ldots,8}$ belongs to the odd parity sector. Particularly, for the asymptotic case $L \gg \xi_{1,2}$ in which the MF e\-ner\-gies become negligible, i.e., ${\epsilon_1} = {\epsilon_2} = \epsilon \approx 0$, we thus have that ${\xi_+} = {\xi_-} = {\epsilon_d}/2$. As a consequence, the energy spectrum will collapse into the two energy levels ${\epsilon_d}/2 \pm \sqrt{ {({{\epsilon}_d}/2)^2} + {{\lambda}^2_1}+ {{\lambda}_2^2} }$ with fourfold degeneracy.
 
For the system of RFs we set the parameter $\widetilde\lambda_j=0$ into Eq.~\eqref{eq:eq000010}. On the one hand, from Eq.~\eqref{eq:eq000011} it follows that $\mathcal{H}_0$ exhibits a one-dimensional matrix block corresponding to the occupation quantum number $N = 0$ and also a three-dimensional block respective to $N = 2$. On the other hand, $\mathcal{H}_1$ in Eq.~\eqref{eq:eq000012} presents a three-dimensional matrix block regarding $N = 1$ and a one-dimensional block for the $N = 3$ parity sector. The set of eigenstates ${\{|{E_j}\rangle\}_{j = 1,\ldots,8}}$ and energies ${\{ {E_j} \}_{j = 1,\ldots,8}}$ of the Hamiltonian for RFs are listed in Table~\ref{tab:tab000002}.


\section{Dynamics of the open quantum system}
\label{sec:sec0003}

In this section we will describe the dynamics of both the systems of MFs$+$QD and RFs$+$QD undergoing the dissipative effects from the coupling to a fermionic reservoir $\mathcal{B}$ that is initialized at the equilibrium state (see Fig.~\ref{fig:fig00001} for details). To do so, we set ${H_S} = {H_{M,R}}$ as the generalized Hamiltonian of the systems [see Eq.~\eqref{eq:eq000008}], which recovers either the Hamiltonian of MFs$+$QD [see Eq.~\eqref{eq:eq000006}] or RFs$+$QD [see Eq.~\eqref{eq:eq000007}] depending on adjustment of the physical para\-me\-ters. The QD is weakly coupled to an environment of free fermions at finite temperature, with Hamiltonian ${H_{\mathcal{B}}} = {\sum_k}\, {\varepsilon_k}{c_k^{\dagger}}{c_k}$, where ${c_k^{\dagger}}$ (${c_k}$) is the creation (annihilation) operator respective to the $k$-th fermionic mode, while ${\varepsilon_k}$ is the energy~\cite{PhysRevB.101.155134,PhysRevE.102.012136}. The inte\-rac\-tion between the system and the bath is described by the Hamiltonian ${H_I} = {\sum_k} \, {g_k} ({d^{\dagger}}{c_k} + d{c_k^{\dagger}})$, where $g_k$ is the coupling strength. Hence the Hamiltonian of the joint system reads $H := {H_S} + {H_{\mathcal{B}}} + {H_I}$.
\begin{table}[t]
\caption{Eigenstates and energies of the Hamiltonian in Eq.~\eqref{eq:eq000010} for the case of regular fermions (${\widetilde{\lambda}_j} = 0$). Here we have defined the parameters $\xi_{\pm}=(\epsilon_d\pm\epsilon)/2$ and ${\Delta_{\pm}} = \sqrt{{\xi_{\pm}^2} + {\lambda^2_1}+\lambda_2^2}$, while ${b_{\mu\nu}^{-1}} ={\sqrt{\, 2{\Delta_{\mu}}({\Delta_{\mu}} + \nu{\xi_{\mu}})}}$ and ${c_{\mu\nu}} = ({\xi_{\mu}} + \nu{\Delta_{\mu}}){b_{\mu\nu}}$, with $\mu, \nu=\{+,- \}$.}
\begin{center}
\begin{tabular}{ll}
\hline\hline
Eigenstate & Energy  \\
\hline
$|{E_1}\rangle = \ket{\vac}$ & ${E_1} = - \epsilon$ \\
$|{E_2}\rangle = \frac{1}{\sqrt{ {\lambda_1^2} + {\lambda_2^2}}}\, ({\lambda_1}{f_2^{\dagger}} - {\lambda_2}{f_1^{\dagger}})\ket{\vac}$ & ${E_2} = 0$  \\
$|{E_3}\rangle = \frac{1}{\sqrt{ {\lambda_1^2} + {\lambda_2^2}}}\, ({\lambda_1}{f_1^\dag} + {\lambda_2}{f_2^{\dagger}}){d^\dag}\ket{\vac}$ & ${E_3} = {\epsilon_d}$  \\
$|{E_4}\rangle = {f_1^{\dagger}}{f^\dag_2}{d^\dag}\ket{\vac} $ & ${E_4} = {\epsilon_d} + \epsilon$  \\
$|{E_5}\rangle = \left({b_{--}}({\lambda_1}{f^\dag_1} + {\lambda_2}{f^\dag_2}) + {c_{--}}{d^\dag}\right)\ket{\vac}$ & ${E_5} = {\xi_-} - {\Delta_-}$ \\
$|{E_6}\rangle = \left({b_{-+}}({\lambda_1}{f^\dag_2} - {\lambda_2}{f^\dag_1}){d^\dag} + {c_{-+}}{f_1^{\dagger}}{f_2^{\dagger}}\right)\ket{\vac}$ & ${E_6} = {\xi_+} - {\Delta_-}$ \\
$|{E_7}\rangle = \left({b_{-+}}({\lambda_1} {f^\dag_1} + {\lambda_2}{f^\dag_2}) + {c_{-+}}{d^\dag}\right)\ket{\vac}$ & ${E_7} = {\xi_-} + {\Delta_-}$ \\
$|{E_8}\rangle = \left({b_{--}}({\lambda_2}{f^\dag_1} - {\lambda_1}{f^\dag_2}){d^{\dagger}} - {c_{--}}{f^\dag_1}{f^\dag_2}\right)\ket{\vac}$ & ${E_8} = {\xi_+} + {\Delta_-}$ \\
\hline\hline
\end{tabular}
\label{tab:tab000002}
\end{center}
\end{table}

\subsection{Density matrix formalism}

To obtain the dynamics of the physical quantities of the system, we employ the well-known density-matrix formalism, within which the dynamics of the system can be obtained by tracing out the environmental degrees of freedom. The re\-sul\-ting quantum master equation can be written as~\cite{PhysRevA.60.91,RevModPhys.89.015001}
\begin{align}
\label{eq:eq000013}
\frac{d{\rho_S}(t)}{dt} &= - i[{H_S} ,{\rho_S} (t) ] + {\int_0^t} d\tau  {{\alpha}^+}(t - \tau) [({V_{\tau - t}} \, {d^{\dagger}}) {\rho_S} (t) ,d] \nonumber\\
&+{\int_0^t} d\tau {{\alpha}^-}(t - \tau)[({V_{\tau - t}} \, d) \, {\rho_S} (t), {d^{\dagger}}] + \text{H.c.} ~,
\end{align}
in which ${V_{\tau - t}}\bullet = {e^{ i (\tau - t) {H_S}}} \bullet \, {e^{- i (\tau - t) {H_S}}}$, and the correlation functions read
\begin{equation}
\label{eq:eq000014}
{\alpha^+} (t) = {\int_0^{\infty}} d\omega J(\omega) {N_F}(\omega){e^{i\omega t}} ~, 
\end{equation}
and
\begin{equation}
\label{eq:eq000015}
{\alpha^-} (t) = {\int_0^{\infty}}  d\omega J(\omega)({N_F}(\omega) + 1){e^{-i\omega t}} ~.
\end{equation}
Here, ${N_F}({\omega}) = {{[\exp(\beta{\omega})+1]}^{-1}}$ is the Fermi-Dirac distribution describing the fermionic reservoir at a given temperature $T= ({k_B}\beta)^{-1}$, and $J(\omega) = {g^2}(\omega)\, {\left|{ {\partial {\omega}(k)}/{\partial k}} \right|}^{-1}$ is the spectral density of the environment, in which $g(\omega)$ is the density of states of the bath. Hereafter we set Boltzmann's and Planck's constants to the unity, i.e., $k_B = \hbar = 1$. Furthermore, we have implicitly assumed the chemical potential of the bath to be zero.

\begin{figure}[t]
\begin{center}
\includegraphics[scale=0.9]{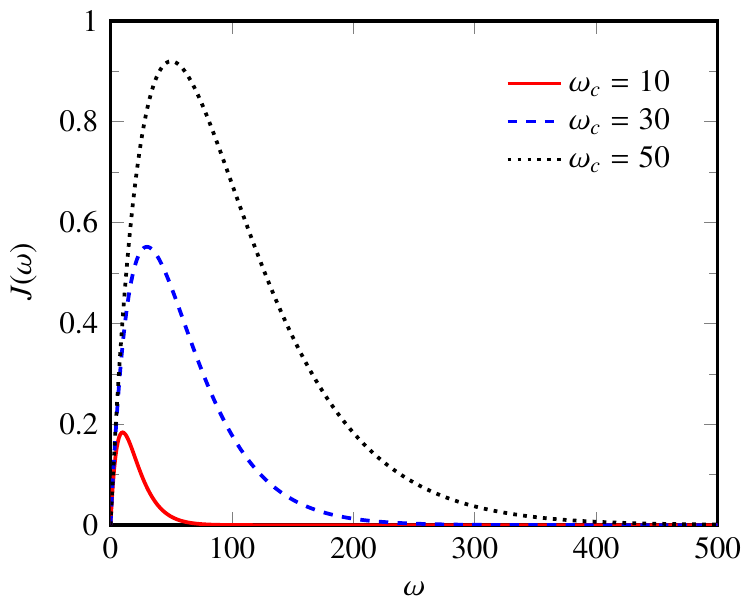}
\caption{(Color online) Spectral density $J(\omega)$ for $s = 1$ ({\it ohmic} case), coupling strength $\gamma = 0.05$, and cutoff frequencies ${\omega_c} = 10$ (red solid line), ${\omega_c} = 30$ (blue dashed line), and ${\omega_c} = 50$ (black dotted line).}
\label{fig:fig00002}
\end{center}
\end{figure}
%
The behavior of the system can be described by considering an accurate model for the spectral density at low frequencies. From now on we will focus on the effect of a generic bath on the system as described by the spectral density~\cite{RevModPhys.59.1,CALDEIRA1983587,Weiss_2008_book}
\begin{equation}
\label{eq:eq000016}
J(\omega) = \gamma \, {{\omega}^s} \, {{\omega}_c^{1 - s}} {e^{-\omega/{{\omega}_c}}} ~,
\end{equation}
for all $s > 0$, where $\gamma$ is the coupling strength of the system and the environment. The environment can be classified as \textit{subohmic} ($0 < s < 1$), \textit{ohmic} ($s=1$), and \textit{superohmic} ($s>1$)~\cite{CALDEIRA1983587,RevModPhys.59.1,Weiss_2008_book}. The exponential factor in Eq.~\eqref{eq:eq000016} provides a smooth cut-off for the spectral density, which is modulated by the frequency $\omega_c$. The frequency $\omega_c$ describes the decaying time scale of the environment as ${\tau_c}\sim {\omega_c^{-1}}$. The Markov limit would correspond to the case ${{\tau}_c} \ll {1/{\Gamma}}$, i.e., when the correlation time $\tau_c$ is much smaller than the typical dissipation time scale of the system given by $\Gamma \sim {\int_0^{\infty}} dx\, {{\alpha}^-}(t - x)$. In Fig.~\ref{fig:fig00002}, we show the spectral density $J(\omega)$ in the ohmic case $s = 1$, for ${\omega_c} = 10$ (non-Markovian) and ${\omega_c} = 50$ (Markovian), also setting $\gamma = 0.05$. Generically speaking, the larger $\omega_c$ the more Markovian is the dynamics of the system.

To solve the master equation, one may recast the marginal state ${{\rho}_S}(t)$ in terms of the occupation number basis $\{|{n_1},{n_2},{n_d}\rangle\}$ as
\begin{equation}
\label{eq:eq000017}
{{\rho}_S}(t) = {\sum_{ \mathbf{k}, \mathbf{m} }} \, {{A}^{ {k_1}, \, {k_2}, \, {k_d}}_{ {m_1}, \, {m_2},\, {m_d} }}(t)\,  |{k_1},{k_2},{k_d}\rangle\langle{m_1},{m_2},{m_d}| ~,
\end{equation}
where $\mathbf{k} = ({k_1},{k_2},{k_d} )$, $\mathbf{m} = ( {m_1},{m_2},{m_d})$, with ${k_j} = \{0,1\}$, and ${m_j} = \{0,1\}$ for $j = \{1,2,d\}$, while
\begin{equation}
\label{eq:eq000018}
{{A}^{ {k_1},\,{k_2},\, {k_d}}_{ {m_1},\, {m_2},\, {m_d} }}(t) = \langle{k_1},{k_2},{k_d}|{{\rho}_S}(t)|{m_1},{m_2},{m_d}\rangle ~.
\end{equation}
Plugging Eq.~\eqref{eq:eq000018} into Eq.~\eqref{eq:eq000013}, we obtain a set of coupled differential equations for the time-dependent coefficients $\{ {{A}^{ {k_1}, \, {k_2}, \, {k_d}}_{ {m_1}, \, {m_2},\,{m_d} }}(t) \}_{\mathbf{k}, \mathbf{m}}$, whose solution fully characterizes the reduced density matrix ${{\rho}_S}(t)$, for a given initial state ${{\rho}_S}(0)$ of the system. We refer to the Appendix~\ref{sec:appendix0A} for details on simplifying the master equation.


The density matrix ${{\rho}_S}(t)$ stores information about the evolution of the system. The solution of the master equation in Eq.~\eqref{eq:eq000013}, also using Eq.~\eqref{eq:eq000017}, allows us to study the occupation numbers $\{ \langle{\hat{n}_{\mu}}\rangle\}_{\mu = 1,2,d}$ of the QD and the two fermions, which can be written as 
\begin{equation}
\label{eq:eq000019}
\langle{\hat{n}_{\mu}}\rangle = {\sum_{{n_1},\,{n_2},\, {n_d}}}\, {n_{\mu}} \, {{A}^{ {n_1}, \, {n_2}, \, {n_d}}_{ {n_1}, \, {n_2},\, {n_d} }} \, (t) ~,
\end{equation}
where ${\hat{n}_d} = {d^{\dagger}}d$, ${\hat{n}_1} = {f_1^{\dagger}}{f_1}$, and ${\hat{n}_2} = {f_2^{\dagger}}{f_2}$ are the fermion number operators.


\subsection{Entanglement and quantum coherences}
\label{sec:sec0004}


\begin{figure}[t!]
\begin{center}
\includegraphics[scale=0.925]{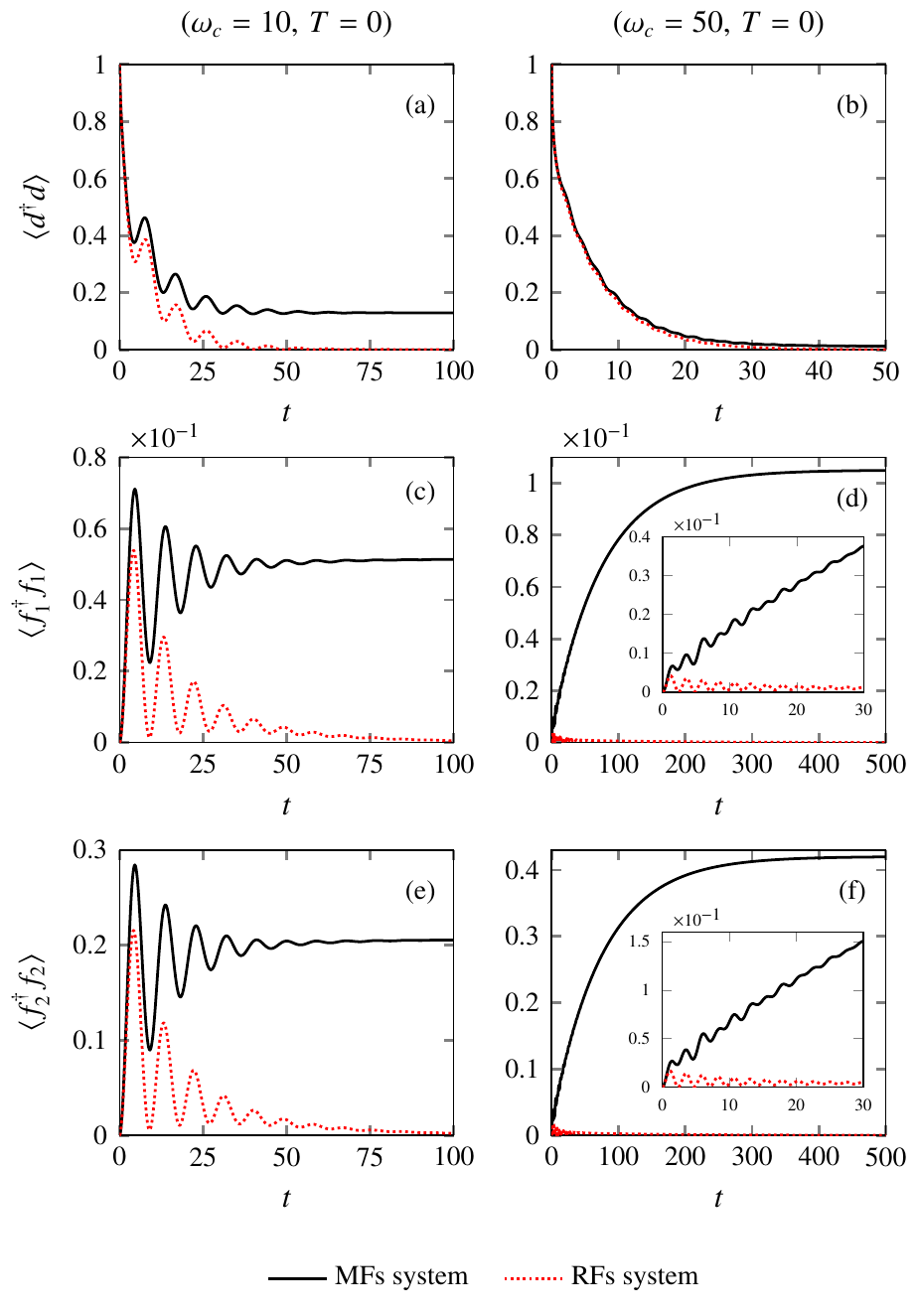}
\caption{(Color online) Populations of the subsystem of MFs$+$QD (black solid line) and RFs$+$QD (red dotted line), where $\langle \bullet \rangle = \text{Tr}(\bullet \, {\rho_S}(t))$, with ${\rho_S}(t)$ being the reduced density matrix obtained from the master equation in Eq.~\eqref{eq:eq000013}. The system of MFs$+$QD (RFs$+$QD) is initialized at the state ${{\rho}_S}(0) = |\tilde{1}\rangle\langle\tilde{1}|$, with $|\tilde{1}\rangle = {d^{\dagger}}\ket{\vac}$, and coupled to a fermionic reservoir at zero temperature $(T = 0)$. Here we set the coupling strength $\gamma = 0.05$, $s = 1$, and cutoff frequencies ${\omega_c} = 10$, non-Markovian regime (left panels), and ${\omega_c} = 50$, Markovian regime (right panels).}
\label{fig:fig00003}
\end{center}
\end{figure}
\begin{figure}[t!]
\begin{center}
\includegraphics[scale=0.925]{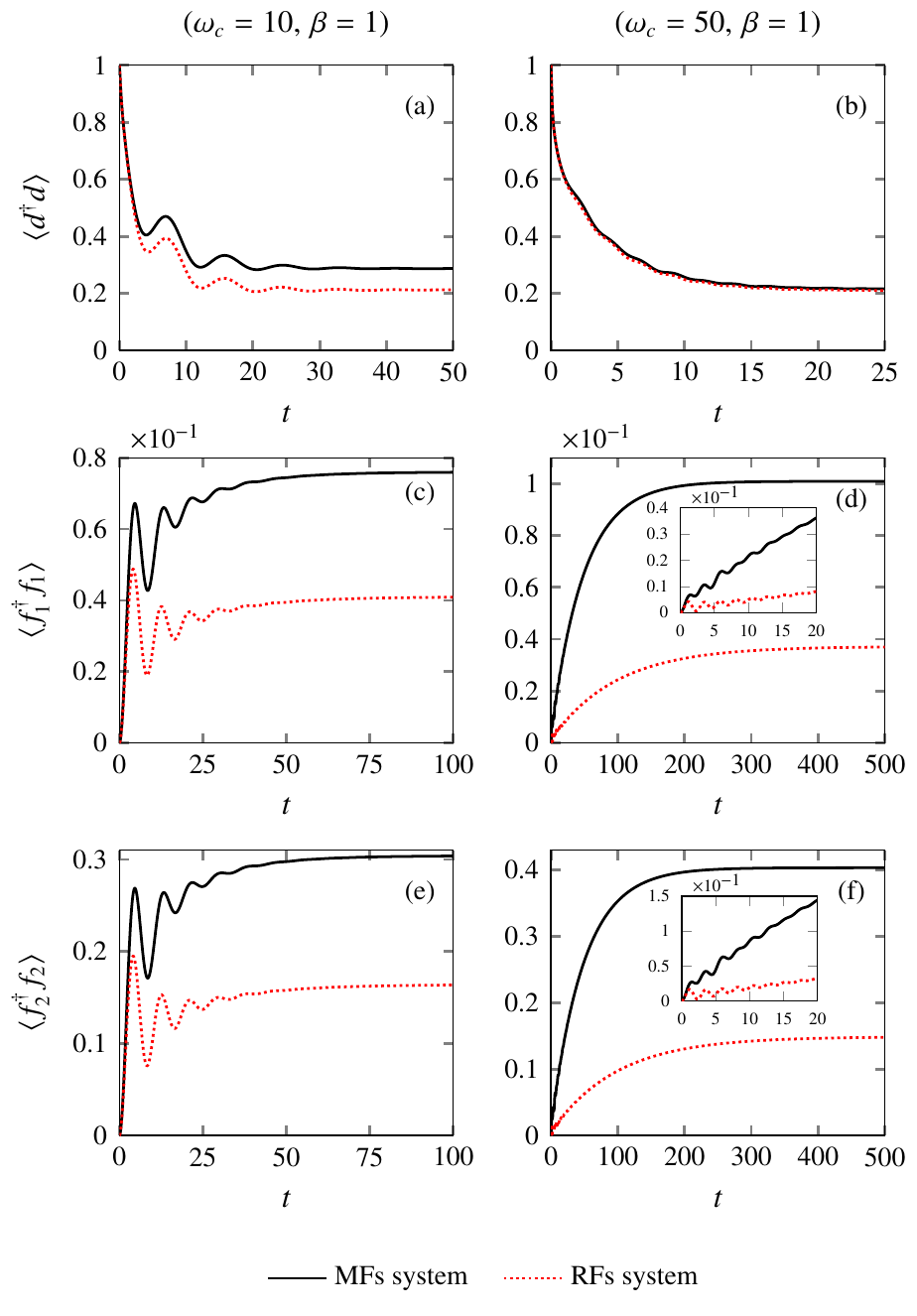}
\caption{(Color online) 
Populations of the subsystem of MFs$+$QD (black solid line) and RFs$+$QD (red dotted line), where $\langle \bullet \rangle = \text{Tr}(\bullet \, {\rho_S}(t))$, with ${\rho_S}(t)$ being the reduced density matrix obtained from the master equation in Eq.~\eqref{eq:eq000013}. The system of MFs$+$QD (RFs$+$QD) is initialized at the pure state ${{\rho}_S}(0) = |\tilde{1}\rangle\langle\tilde{1}|$, with $|\tilde{1}\rangle = {d^{\dagger}}\ket{\vac}$, and coupled to a fermionic reservoir at finite temperature $(\beta = 1)$. Here we set the coupling strength $\gamma = 0.05$, $s = 1$, and cutoff frequencies ${\omega_c} = 10$, non-Markovian regime (left panels), and ${\omega_c} = 50$, Markovian regime (right panels).}
\label{fig:fig00004}
\end{center}
\end{figure}

Here we will introduce the minimal theoretic framework to study the role of entanglement and quantum cohe\-ren\-ces in both the systems of MFs and RFs, with respect to the two-body reduced states
\begin{equation}
\label{eq:eq000020}
{\rho_{jl}}(t) = {\sum_{ {k_j}, \, {k_l} }}~{\sum_{ {m_j}, \, {m_l} }}~{{A}^{ {k_j},\, {k_l}}_{ {m_j},\, {m_l} }}(t) \, |{k_j},{k_l}\rangle\langle{m_j},{m_l}| ~,
\end{equation}
with $j,l = \{1,2,d\}$ and $j \neq l$, ${k_j} = \{0,1\}$, and ${m_j} = \{0,1\}$, where
\begin{equation}
\label{eq:eq000021}
{{A}^{ {k_j},\, {k_l}}_{ {m_j},\, {m_l} }}(t) = {\sum_{\substack{{k_y} : \, y \neq j \neq l }}}~{{A}^{ {k_1},\, {k_2},\, {k_d}}_{ {m_1},\, {m_2},\, {k_d} }}(t) ~.
\end{equation}
For our purposes, we address quantum correlations according to the concurrence, a bipartite entanglement quantifier~\cite{RevModPhys.81.865}, while for quantum coherences our analysis is based on the so-called $\ell_1$ norm of coherence~\cite{RevModPhys.89.041003}. The concurrence is defined as~\cite{PhysRevLett.78.5022,PhysRevLett.80.2245}
\begin{equation}
\label{eq:eq000022}
\text{Conc}[\rho] = \max(0,{\varpi_1} - {\varpi_2} - {\varpi_3} - {\varpi_4}) ~,
\end{equation}
where $\{ {\varpi_j} \}_{j = 1,\ldots,4}$ are the eigenvalues in decreasing order of the matrix 
\begin{equation}
\label{eq:eq000023}
R[\rho] := \sqrt{ \sqrt{\rho}\, ({\sigma_y}\otimes{\sigma_y}){\rho^*}({\sigma_y}\otimes{\sigma_y})\sqrt{\rho} } ~.
\end{equation}
Opposite to entanglement, quantum coherence is a basis dependent quantity, and thus its formulation requires one to fix some preferred basis states. Hereafter we will adopt the re\-fe\-rence basis $\{ | {n_1},{n_2}\rangle, |{n_1},{n_d}\rangle, |{n_2},{n_d}\rangle \}$, with ${n_j} = \{0,1\}$ for $j = \{1,2,d\}$. To characterize the quantum coherence stored in the marginal states in Eq.~\eqref{eq:eq000020}, we consider the so-called ${{\ell}_1}$ norm of coherence, i.e., a monotonic distance-based quantifier of coherence written as~\cite{BCP2014}
\begin{equation}
\label{eq:eq000024}
{\mathcal{C}_{\ell_1}}[{\rho_{jl}}(t)] = {\sum_{\substack{j\neq l ; \, {\mathbf{n}_{jl}} \neq {\mathbf{k}_{jl}}  }}} \, |\langle{k_j},{k_l}|\, {\rho_{jl}}(t) |{n_j},{n_l}\rangle|  ~,
\end{equation}
where ${\mathbf{n}_{jl}} := ({n_j},{n_l} )$ and ${\mathbf{k}_{jl}} := ({k_j},{k_l} )$, with ${n_j} = \{0,1\}$ and ${k_j} = \{0,1\}$, for $j,l = \{1,2,d\}$. In what follows, we present our numerical results obtained from the equations presented above.

\begin{figure*}[!t]
\begin{center}
\includegraphics[scale=0.95]{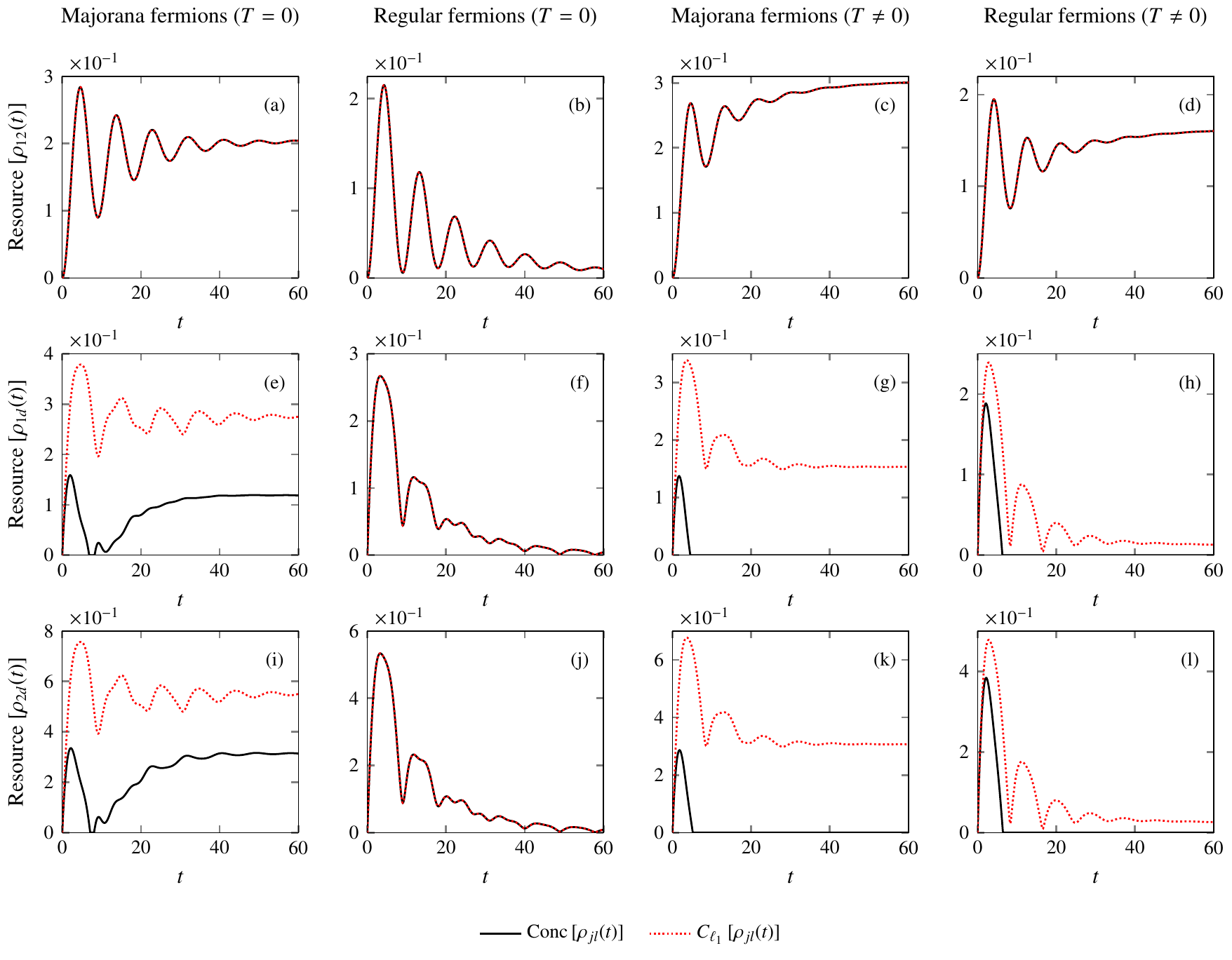}
\caption{(Color online) Comparison between concurrence (black solid line) and $\ell_1$ norm of coherence (red dashed line) for the non-Markovian dynamics (${\omega_c} = 10$) of MFs and RFs. Panels (a),~(b),~(e),~(f),~(i), and~(j) refer to the case of zero temperature ($T = 0$), while panels (c),~(d),~(g),~(h),~(k), and~(l) show the curves for the bath at finite temperature ($\beta = 1$). Here we choose the initial state of the system MFs$+$QD (RFs$+$QD) given by ${{\rho}_S}(0) = |\tilde{1}\rangle\langle\tilde{1}|$, with $|\tilde{1}\rangle = {d^{\dagger}}\ket{\vac}$, and $\gamma = 0.05$, $s = 1$, $\epsilon = 0.5$, ${{\epsilon}_d} = 0.5$, and ${\lambda_2} = 2{\lambda_1} = 0.2$.}
\label{fig:fig00005}
\end{center}
\end{figure*}

\begin{figure*}[!t]
\begin{center}
\includegraphics[scale=0.95]{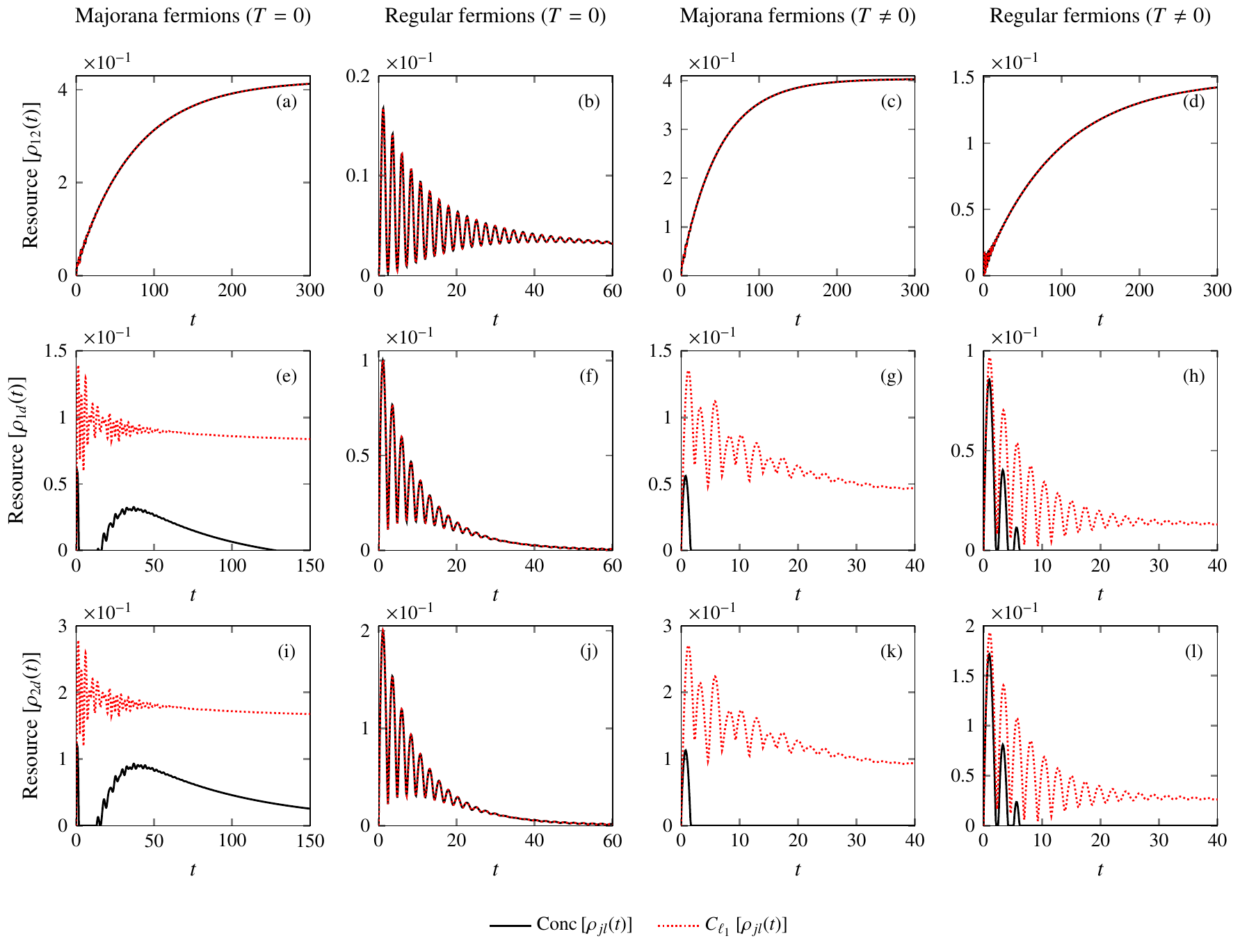}
\caption{(Color online) Comparison between concurrence (black solid line) and $\ell_1$ norm of coherence (red dashed line) for the Markovian dynamics (${\omega_c} = 50$) of MFs and RFs. Panels (a),~(b),~(e),~(f),~(i), and~(j) refer to the case of zero temperature ($T = 0$), while panels (c),~(d),~(g),~(h),~(k), and~(l) show the curves for the bath at finite temperature ($\beta = 1$). Here we choose the initial state of the system MFs$+$QD (RFs$+$QD) given by ${{\rho}_S}(0) = |\tilde{1}\rangle\langle\tilde{1}|$, with $|\tilde{1}\rangle = {d^{\dagger}}\ket{\vac}$, and $\gamma = 0.05$, $s = 1$, $\epsilon = 0.5$, ${{\epsilon}_d} = 0.5$, and ${\lambda_2} = 2{\lambda_1} = 0.2$.}
\label{fig:fig00006}
\end{center}
\end{figure*}

\begin{figure*}[ht]
\begin{center}
\includegraphics[scale=0.95]{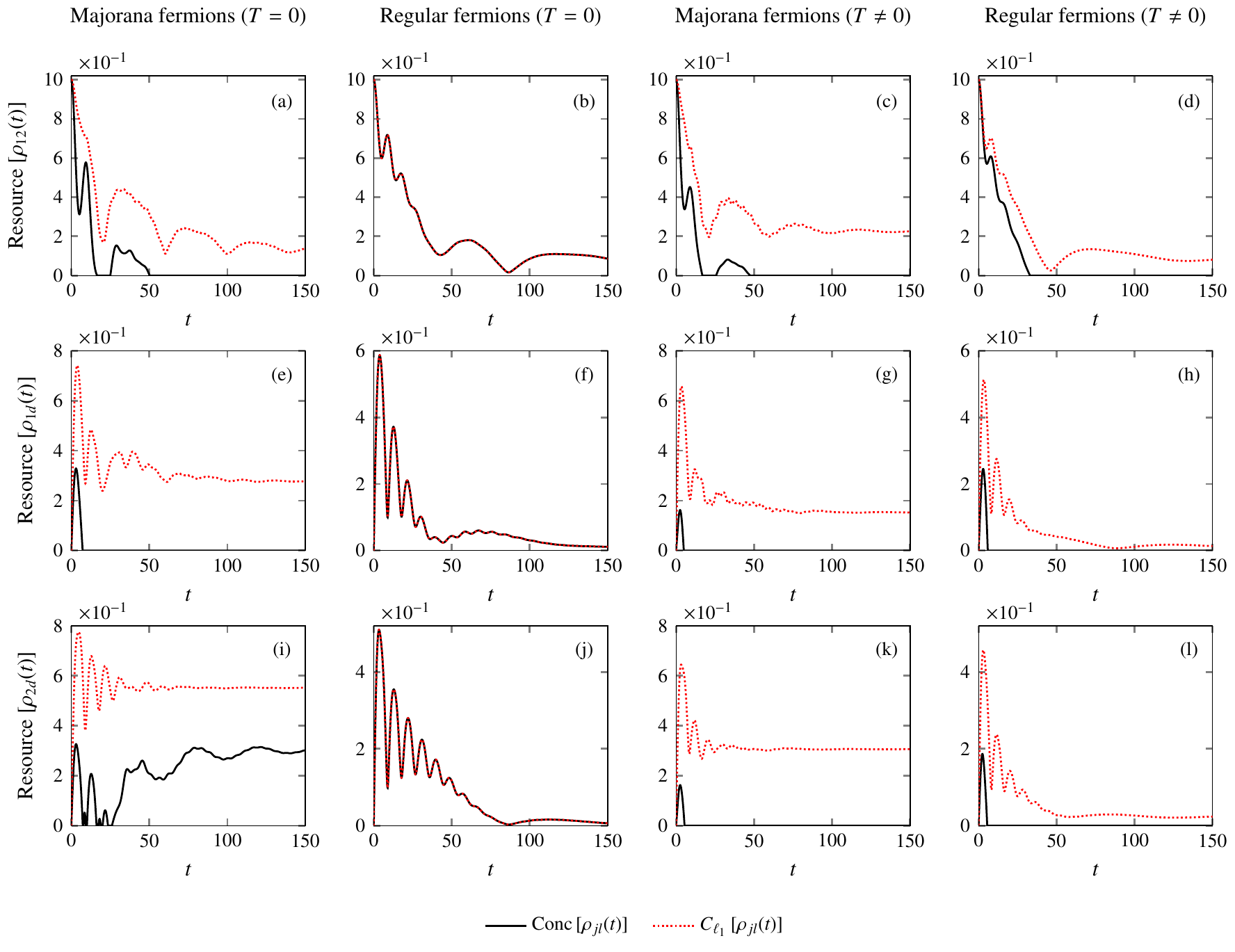}
\caption{(Color online) Comparison between concurrence (black solid line) and $\ell_1$ norm of coherence (red dashed line) for the non-Markovian dynamics (${\omega_c} = 10$) of MFs and RFs. Panels (a),~(b),~(e),~(f),~(i), and~(j) refer to the case of zero temperature ($T = 0$), while panels (c),~(d),~(g),~(h),~(k), and~(l) show the curves for finite temperature ($\beta = 1$). Here we choose the initial state of the system MFs$+$QD (RFs$+$QD) given by ${{\rho}_S}(0) = |\tilde{+}\rangle\langle\tilde{+}|$, with $|\tilde{+}\rangle := \frac{1}{\sqrt{2}}\, ({f_1^{\dagger}} + {f_2^{\dagger}})\ket{\vac}$, and $\gamma = 0.05$, $s = 1$, $\epsilon = 0.5$, ${{\epsilon}_d} = 0.5$, and ${\lambda_2} = 2{\lambda_1} = 0.2$.}
\label{fig:fig00007}
\end{center}
\end{figure*}

\begin{figure*}[!t]
\begin{center}
\includegraphics[scale=0.95]{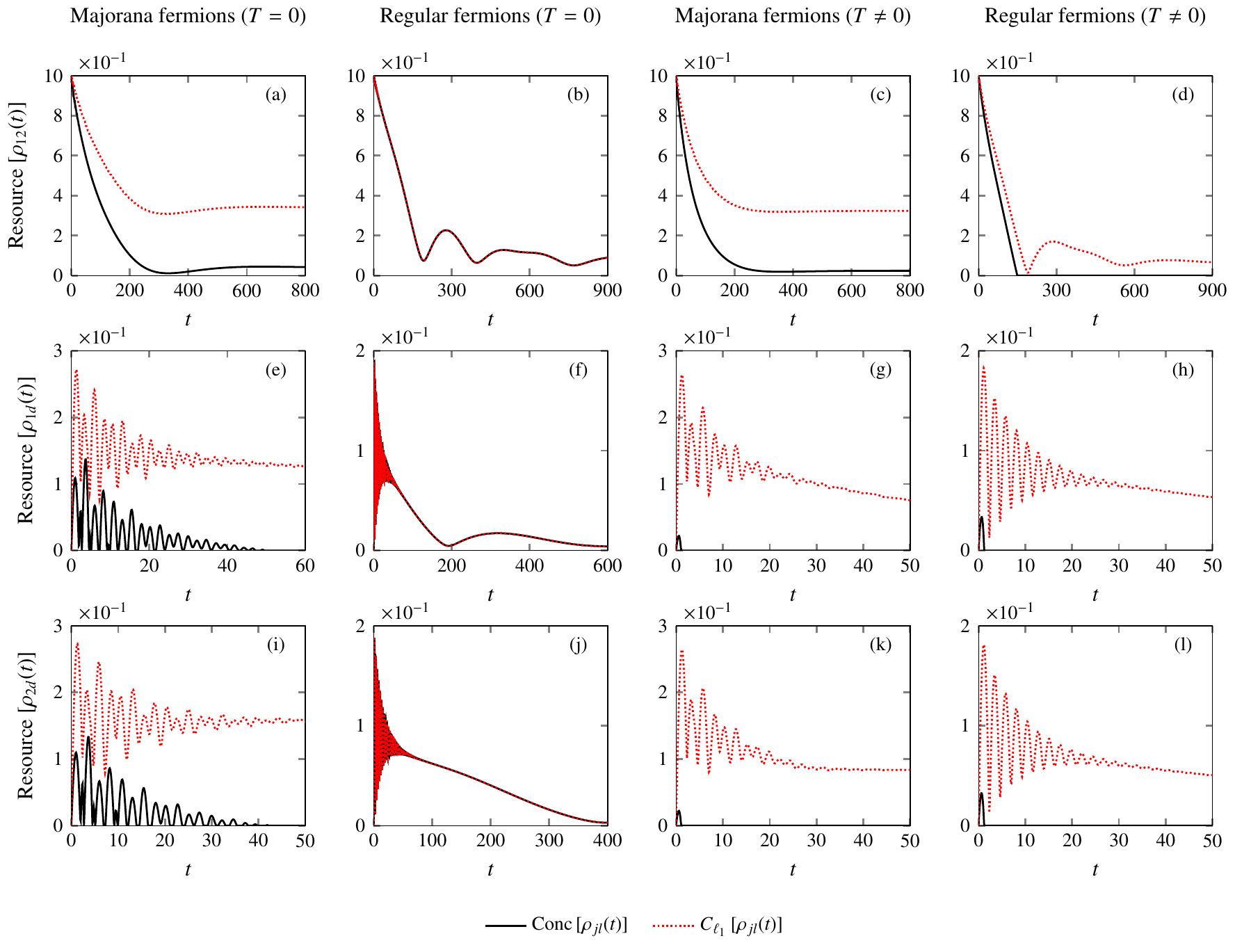}
\caption{(Color online) Comparison between concurrence (black solid line) and $\ell_1$ norm of coherence (red dashed line) for the Markovian dynamics (${\omega_c} = 50$) of MFs and RFs. Panels (a),~(b),~(e),~(f),~(i), and~(j) refer to the case of zero temperature ($T = 0$), while panels (c),~(d),~(g),~(h),~(k), and~(l) show the curves at finite temperature ($\beta = 1$). Here we choose the initial state of the system MFs$+$QD (RFs$+$QD) given by ${{\rho}_S}(0) = |\tilde{+}\rangle\langle\tilde{+}|$, with $|\tilde{+}\rangle := \frac{1}{\sqrt{2}}\, ({f_1^{\dagger}} + {f_2^{\dagger}})\ket{\vac}$, and $\gamma = 0.05$, $s = 1$, $\epsilon = 0.5$, ${{\epsilon}_d} = 0.5$, and ${\lambda_2} = 2{\lambda_1} = 0.2$.}
\label{fig:fig00008}
\end{center}
\end{figure*}

\begin{figure*}[!t]
\begin{center}
\includegraphics[scale=0.95]{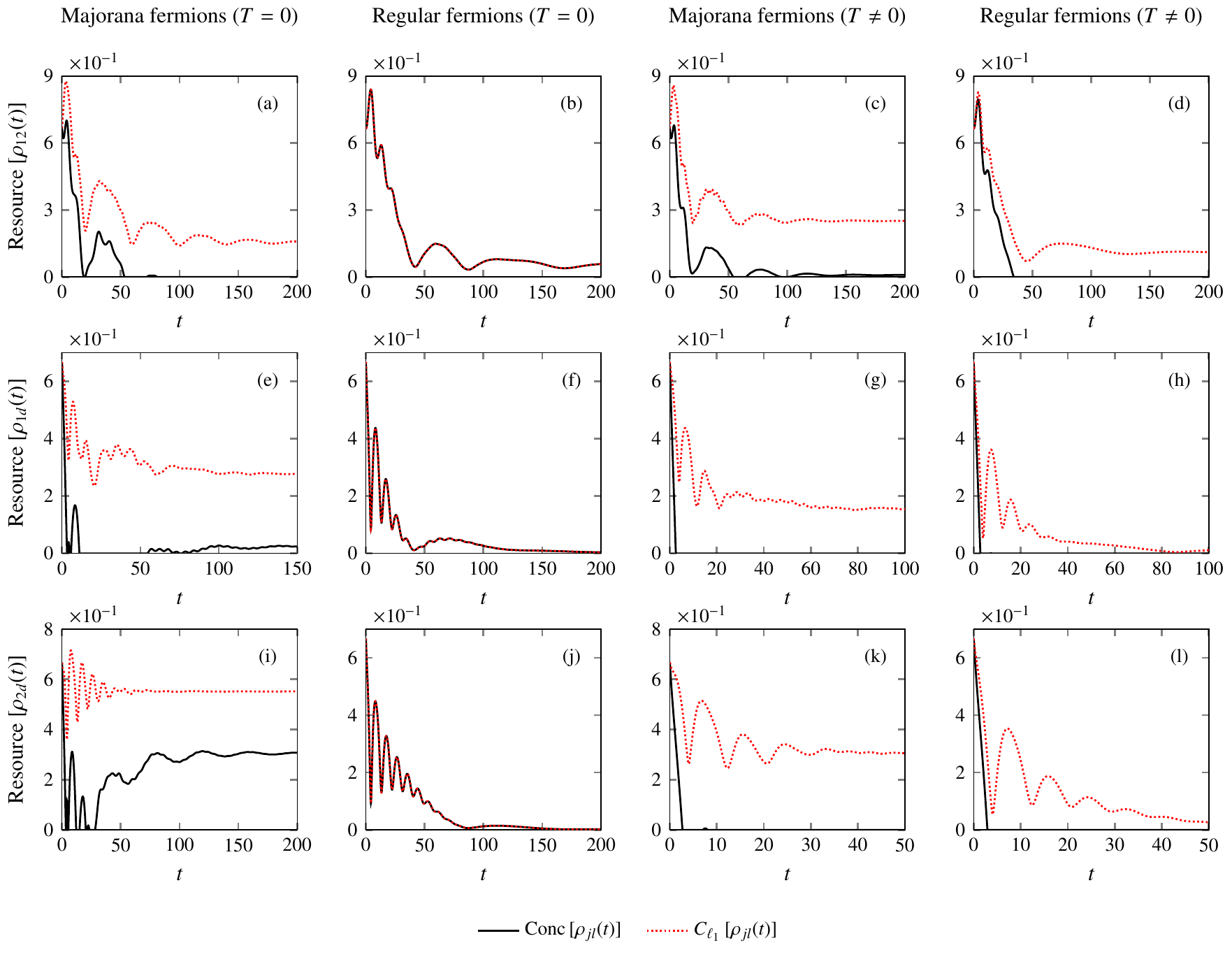}
\caption{(Color online) Comparison between concurrence (black solid line) and $\ell_1$ norm of coherence (red dashed line) for the non-Markovian dynamics (${\omega_c} = 10$) of MFs and RFs. Panels (a),~(b),~(e),~(f),~(i), and~(j) refer to the case of zero temperature ($T = 0$), while panels (c),~(d),~(g),~(h),~(k), and~(l) show the curves at finite temperature ($\beta = 1$). Here we choose the initial state of the system MFs$+$QD (RFs$+$QD) given by ${{\rho}_S}(0) = |{W}\rangle\langle{W}|$, with $|{W}\rangle := \frac{1}{\sqrt{3}}\, ({d^{\dagger}} + {f_1^{\dagger}} + {f_2^{\dagger}})\ket{\vac}$, and $\gamma = 0.05$, $s = 1$, $\epsilon = 0.5$, ${{\epsilon}_d} = 0.5$, and ${\lambda_2} = 2{\lambda_1} = 0.2$.}
\label{fig:fig00009}
\end{center}
\end{figure*}

\begin{figure*}[!t]
\begin{center}
\includegraphics[scale=0.95]{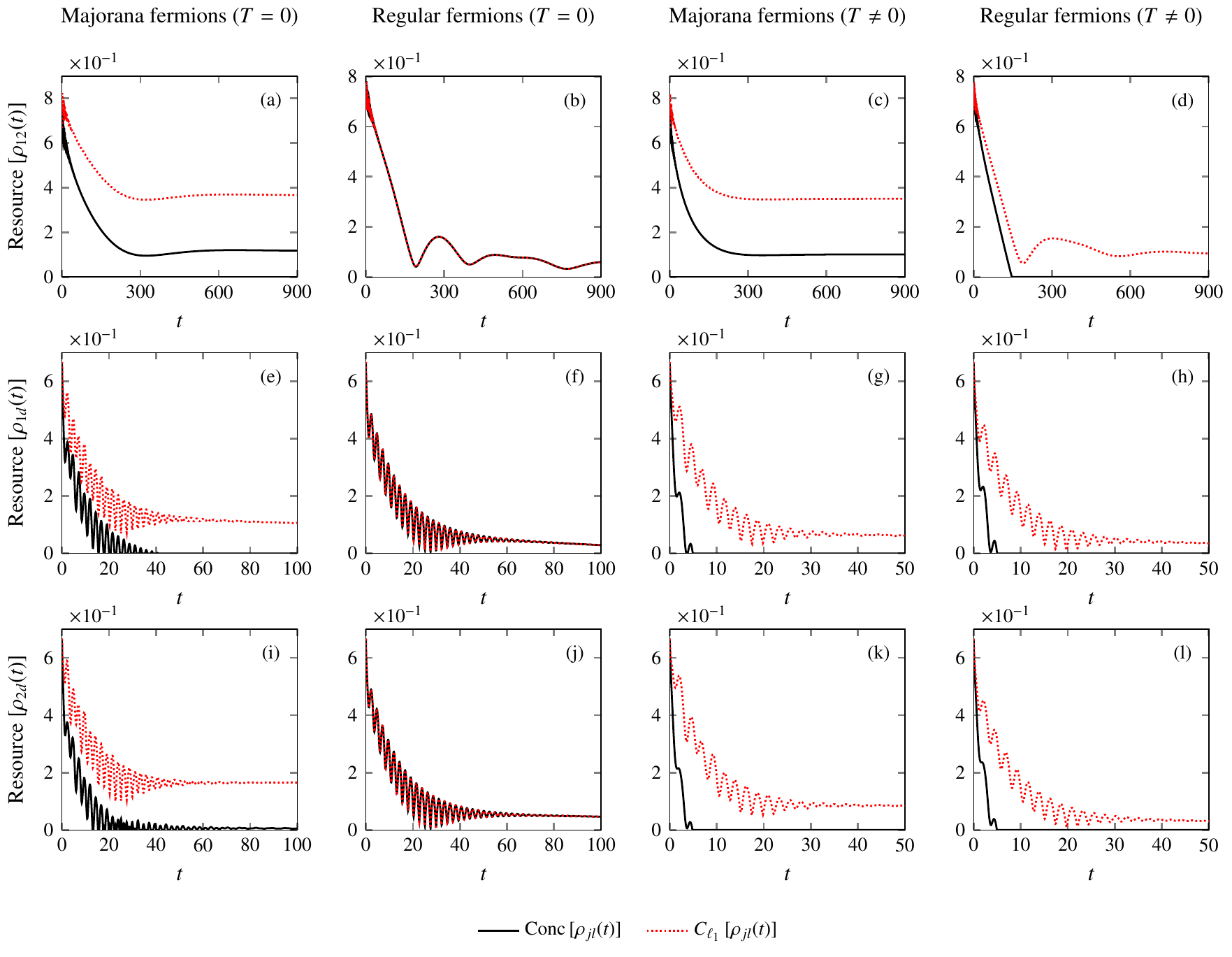}
\caption{(Color online) Comparison between concurrence (black solid line) and $\ell_1$ norm of coherence (red dashed line) for the Markovian dynamics (${\omega_c} = 50$) of MFs and RFs. Panels (a),~(b),~(e),~(f),~(i), and~(j) refer to the case of zero temperature ($T = 0$), while panels (c),~(d),~(g),~(h),~(k), and~(l) show the curves at finite temperature ($\beta = 1$). Here we choose the initial state of the system MFs$+$QD (RFs$+$QD) given by ${{\rho}_S}(0) = |{W}\rangle\langle{W}|$, with $|{W}\rangle := \frac{1}{\sqrt{3}}\, ({d^{\dagger}} + {f_1^{\dagger}} + {f_2^{\dagger}})\ket{\vac}$, and $\gamma = 0.05$, $s = 1$, $\epsilon = 0.5$, ${{\epsilon}_d} = 0.5$, and ${\lambda_2} = 2{\lambda_1} = 0.2$.}
\label{fig:fig00010}
\end{center}
\end{figure*}

\begin{figure*}[t]
\begin{center}
\includegraphics[scale=0.95]{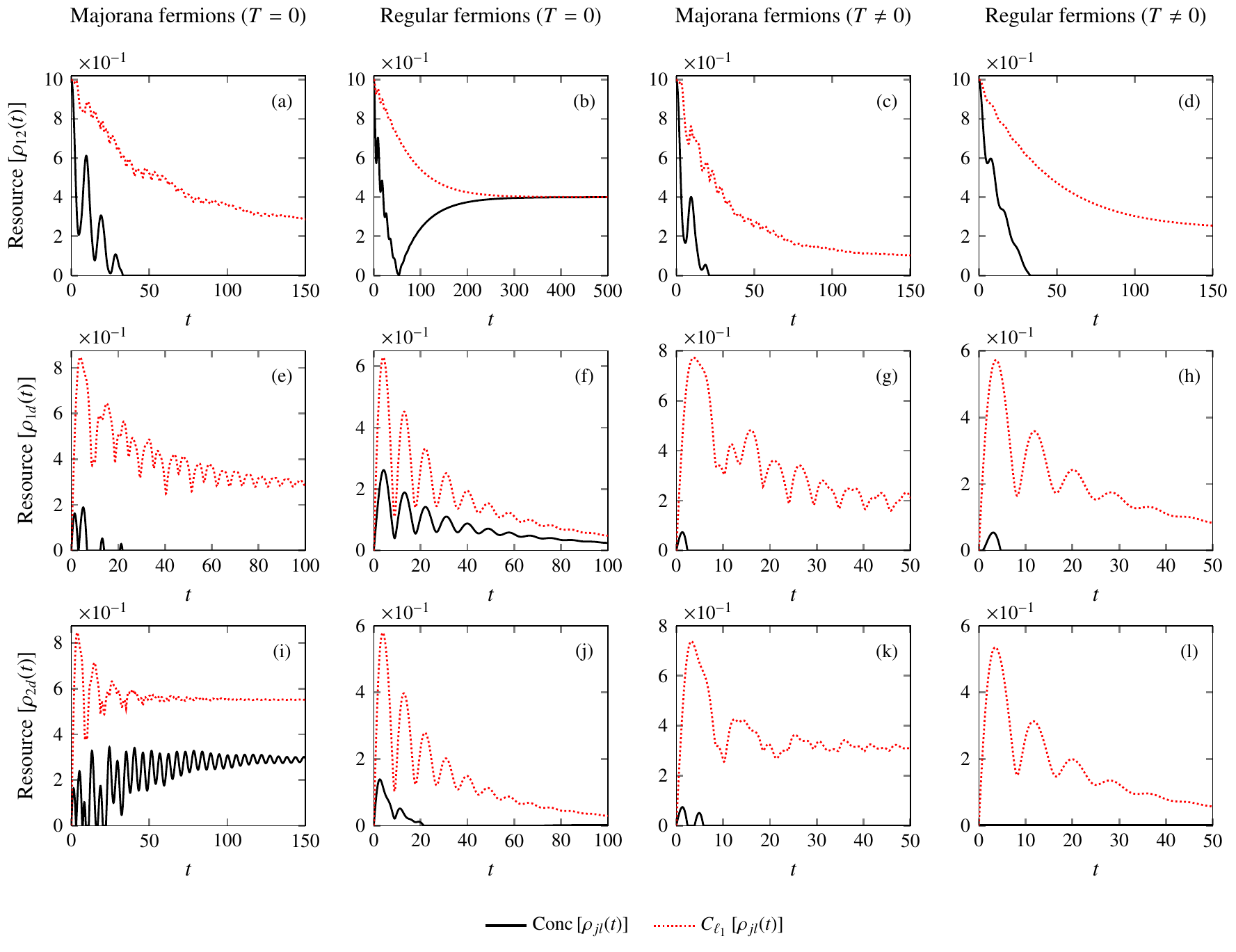}
\caption{(Color online) Comparison between concurrence (black solid line) and $\ell_1$ norm of coherence (red dashed line) for the non-Markovian dynamics (${\omega_c} = 10$) of MFs and RFs at zero ($T = 0$) and finite ($\beta = 1$) temperatures. Here we choose the initial state of the system MFs$+$QD (RFs$+$QD) given by ${{\rho}_S}(0) = |\tilde{\phi}\rangle\langle\tilde{\phi}|$, with $|\tilde{\phi}\rangle := \frac{1}{\sqrt{2}}(\mathbb{I} + {f_1^{\dagger}}{f_2^{\dagger}})\ket{\vac}$, and $\gamma = 0.05$, $s = 1$, $\epsilon = 0.5$, ${{\epsilon}_d} = 0.5$, and ${\lambda_2} = 2{\lambda_1} = 0.2$.}
\label{fig:fig00011}
\end{center}
\end{figure*}

\begin{figure*}[!ht]
\begin{center}
\includegraphics[scale=0.95]{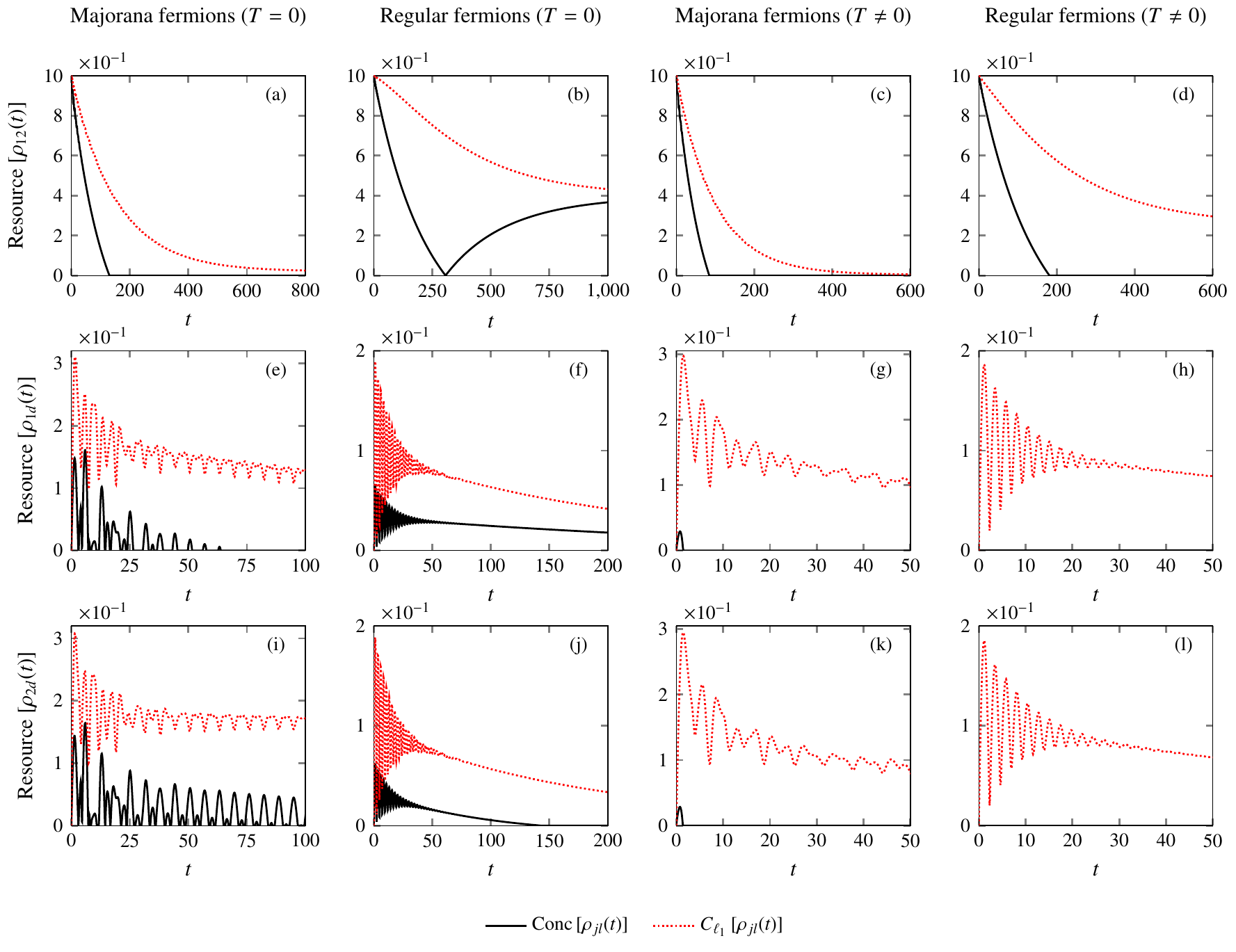}
\caption{(Color online) Comparison between concurrence (black solid line) and $\ell_1$ norm of coherence (red dashed line) for the Markovian dynamics (${\omega_c} = 50$) of MFs and RFs at zero ($T = 0$) and finite ($\beta = 1$) temperatures. Here we choose the initial state of the system MFs$+$QD (RFs$+$QD) given by ${{\rho}_S}(0) = |\tilde{\phi}\rangle\langle\tilde{\phi}|$, with $|\tilde{\phi}\rangle := \frac{1}{\sqrt{2}}(\mathbb{I} + {f_1^{\dagger}}{f_2^{\dagger}})\ket{\vac}$, and $\gamma = 0.05$, $s = 1$, $\epsilon = 0.5$, ${{\epsilon}_d} = 0.5$, and ${\lambda_2} = 2{\lambda_1} = 0.2$.}
\label{fig:fig00012}
\end{center}
\end{figure*}

\section{Numerical results}
\label{sec:sec0004n}

To obtain our numerical results we set the coupling strength $\gamma = 0.05$, ${\lambda_2} = 2{\lambda_1} = 0.2$, and $\epsilon_d=0.5$. Bearing in mind that the chemical potential of the bath is zero, this fixed value of $\epsilon_d$ serves as an energy reference for our calculations. Otherwise stated, we will also use $\epsilon=0.5$. Moreover, we will assume an ohmic bath characterized by $s=1$. Our analysis will consider either Markovian or non-Markovian baths, for which we use ${\omega_c} = 10$ and  ${\omega_c} = 50$, respectively. The results will also be shown for zero temperature ($T=0$) and finite temperature ($\beta = 1$). We will consider initial states within the Hilbert subspace of the system characterized by occupation $N=1$ (odd) and $N=2$ (even).



\subsection{Dynamics of occupations of the single-fermion initial state}
\label{sec:sec0004a}

As we have discussed already, we are mainly interested in the dynamics of quantum coherences and correlations. Before doing so, we will briefly discuss how the occupations evolve in time for a simple case of $N=1$. For this, we consider the initial state ${{\rho}_S}(0) = |\tilde{1}\rangle\langle\tilde{1}|$, with $|\tilde{1}\rangle = {d^{\dagger}}\ket{\vac}$ representing the nonzero fermionic occupation in the QD at time $t = 0$.

The dynamics of the populations $\langle {f^{\dagger}_1} {f_1}\rangle$, $\langle {f^{\dagger}_2} {f_2}\rangle$, and $\langle {d^{\dagger}} d\rangle$ in both MFs$+$QD (black solid line) and RFs$+$QD (red dotted line) setups are shown in Figs.~\ref{fig:fig00003} ($T = 0$) and~\ref{fig:fig00004} ($\beta=1$). Left and right panels refer to cutoff frequencies ${\omega_c} = 10$ (non-Markovian), and ${\omega_c} = 50$ (Markovian), respectively. For $T = 0$, the non-Markovian dynamics ($\omega_c = 10$) of the average occupation $\langle{d^{\dagger}}{d}\rangle$ of the QD decreases and exhibits damped oscillations in both the systems of RFs$+$QD and MFs$+$QD, going to zero in the former case, while the latter reaches a nonzero stationary value [see Fig.~\ref{fig:fig00003}(a)]. The long-time behavior of the occupation vanishes for the RFs$+$QD system because all the levels are above the chemical potential of the system, in which case the fermion leaks into the bath. In contrast, the occupations $\langle{f_j^{\dagger}}{f_j}\rangle$ will grow and oscillate with damped amplitudes, thus approaching a sta\-tio\-na\-ry value which is asymptotically zero for RFs and nonzero for MFs [see Figs.~\ref{fig:fig00003}(c), and~\ref{fig:fig00003}(e)]. Here, the finite occupation of the MFs in the long-time regime results from the terms proportional to $\widetilde\lambda_j$ in the Hamiltonian [see Eq.~\eqref{eq:eq000008}] acting as a source of charge for the system. 

At finite temperature ($\beta = 1$), the occupations will behave quite similarly to the case of zero temperature, except that populations of the QD and both fermions will asymptotically converge to nonzero values at later times of the dynamics [see  Figs.~\ref{fig:fig00004}(a), ~\ref{fig:fig00004}(c), and~\ref{fig:fig00004}(e)]. Here, the occupation is mainly provided by thermal excitation, since $T=1$ is of the order of the energy of the levels $\epsilon=\epsilon_d=0.5$ in this case.

In the Markovian dynamics of the occupations ($\omega_c = 50$) shown in the right panels of Figs.~\ref{fig:fig00003} and~\ref{fig:fig00004}, the population of the QD decreases quite smoothly vanishing at large $t$ for zero tem\-pe\-ra\-ture [see Fig.~\ref{fig:fig00003}(b)], while it reaches a stationary nonzero value for finite temperature ($\beta = 1$) [Fig.~\ref{fig:fig00004}(b)]. Interestingly, the occupation of the QD for both the systems of MFs and RFs show similar behaviors, with a negligible difference. This is because the temperature used is large enough to smooth out any effect of the nonconserving term of the Hamiltonian into the QD occupation. Moreover, the occupations $\langle{f_1^{\dagger}}{f_1}\rangle$ and $\langle{f_2^{\dagger}}{f_2}\rangle$ exhibit nonzero values that increase and reach stationary values for a wide time window in the system of MFs, for both temperatures [see Figs.~\ref{fig:fig00003}(d),~\ref{fig:fig00003}(f),~\ref{fig:fig00004}(d), and~\ref{fig:fig00004}(f)]. However, the occupations of RFs increase initially, but invert this behavior vanishing eventually at later time for $T = 0$ [see Figs.~\ref{fig:fig00003}(d) and~\ref{fig:fig00003}(f)].  At finite temperature  ($\beta = 1$)  these occupations saturate at a nonzero value for longer times [see Figs.~\ref{fig:fig00004}(d) and~\ref{fig:fig00004}(f)]. The insets clearly show fluctuations in the amplitudes of the fermionic occupations at earlier times, which in turn are smoothly suppressed as $t$ increases.

\subsection{Dynamics of coherence and correlations}
\label{sec:sec0004b}

Let us now turn our attention to the numerical analysis for entanglement and quantum coherence for some paradigmatic initial pure states for the systems of MFs$+$QD and RFs$+$QD. Overall, such probe states are representative in the context of quantum information science, with applications ranging from quantum metrology~\cite{T_th_2014,RevModPhys.90.035005} to quantum communication~\cite{Duan_Nature_2001}. These input states store a given amount of coherence and entanglement, i.e., quantum resources that would be sensitive to the dissipative effects from the coupling between the system MFs$+$QD (RFs$+$QD) and the reservoir. Here we will address the robustness of quantum resources for the set of initial states and the interconvertibility of the quantum resources, i.e., verify the conditions where quantum coherence can be consumed and converted into entanglement, and vice versa, under the open quantum dynamics. This is somehow related to the concept of genuine and distributed correlated coherence, which in turn witnesses the amount of quantum coherence that is contained within the correlations of the tripartite system~\cite{PhysRevA.94.022329,JMathPhys_51_414013}.
%

\subsubsection{Single-fermion initial state}

\paragraph{Separable initial state ---} Similar to the previous section, here we consider the initial state  ${{\rho}_S}(0) = |\tilde{1}\rangle\langle\tilde{1}|$, with $|\tilde{1}\rangle = {d^{\dagger}}\ket{\vac}$. Importantly, this initial state is completely uncorrelated, and also incoherent with respect to the reference basis of states $\{ |{n_1},{n_2},{n_d}\rangle\}$. In Figs.~\ref{fig:fig00005} and~\ref{fig:fig00006}, we compare the concurrence and the $\ell_1$ norm of coherence in both the systems MFs$+$QD and RFs$+$QD, for cutoff frequencies ${\omega_c} = 10$ and ${\omega_c} = 50$, respectively. 

In Fig.~\ref{fig:fig00005}, note that the reduced density matrix ${\rho_{12}}(t)$ [see Eq.~\eqref{eq:eq000020}] exhibits nonzero values for concurrence and $\ell_1$ norm of coherence, which in turn coincide in both systems of MFs and RFs, regardless  the temperature [see Figs.~\ref{fig:fig00005}(a),~\ref{fig:fig00005}(b),~\ref{fig:fig00005}(c), and~\ref{fig:fig00005}(d)]. In contrast, for marginal states ${\rho_{1d}}(t)$ and ${\rho_{2d}}(t)$ [see Eq.~\eqref{eq:eq000020}], the dynamics of entanglement and quantum co\-he\-ren\-ce will coincide only in the system of RFs, at zero tem\-pe\-ra\-ture [see Fig.~\ref{fig:fig00005}(f) and~\ref{fig:fig00005}(j)], thus behaving quite diffe\-ren\-tly in the other scenarios. Indeed, at finite temperature ($\beta = 1$), note that concurrence starts growing but  goes suddenly to zero for MFs and RFs, while the quantum coherence exhibits oscillations that are suppressed until approaching a nonzero constant value [see Figs.~\ref{fig:fig00005}(g),~\ref{fig:fig00005}(h),~\ref{fig:fig00005}(k), and~\ref{fig:fig00005}(l)]. In Figs.~\ref{fig:fig00005}(e) and~\ref{fig:fig00005}(i), concurrence exhibits a revival after suddenly going to zero, and then saturates at a constant value for longer times of the dynamics.

Figures~\ref{fig:fig00006}(a),~\ref{fig:fig00006}(b),~\ref{fig:fig00006}(c), and~\ref{fig:fig00006}(d) show that concurrence and quantum coherence of the marginal state ${\rho_{12}}(t)$ coincide for both species of MFs and RFs, regardless the temperature, for the Markovian frequency $\omega_c = 50$. However, for states ${\rho_{1d}}(t)$ and ${\rho_{2d}}(t)$, entanglement and quantum co\-he\-ren\-ce will coincide exclusively for RFs, at zero tem\-pe\-ra\-ture, also exhi\-bi\-ting a highly oscillating behavior [see Figs.~\ref{fig:fig00006}(f) and~\ref{fig:fig00006}(j)]. Indeed, for finite temperature ($\beta = 1$), note that concurrence starts increasing and dropping to zero suddenly for MFs [see Figs.~\ref{fig:fig00006}(g), and~\ref{fig:fig00006}(k)], while it exhibits revivals whose amplitude mostly decreases in the system of RFs [see Figs.~\ref{fig:fig00006}(h) and~\ref{fig:fig00006}(l)]. In addition, quantum cohe\-ren\-ce exhibits a highly oscillating regime, saturating at a nonzero constant value for longer times [see Figs.~\ref{fig:fig00006}(g),~\ref{fig:fig00006}(h),~\ref{fig:fig00006}(k), and~\ref{fig:fig00006}(l)]. In Figs.~\ref{fig:fig00006}(e) and~\ref{fig:fig00006}(i), concurrence exhibits a revival after suddenly going to zero, also experiencing fluctuations in its amplitude that are suppressed as it starts decreasing, and then approaches zero at later times. We emphasize this behavior is strikingly different from the non-Markovian case ($\omega_c = 10$) in Figs.~\ref{fig:fig00005}(e) and~\ref{fig:fig00005}(i), in which both concurrence and quantum coherence remains nonzero for longer times of the dynamics. An important message we can take from the result from Figs.~\ref{fig:fig00005} and~\ref{fig:fig00006} is that the quantum correlations and coherences depend strongly on the temperature for RFs, but are qualitatively similar for MFs at both zero and finite temperatures.  

\paragraph{Superposition initial state ---} Here we set the input state ${{\rho}_S}(0) = |\tilde{+} \rangle\langle\tilde{+}|$ that exhibits nonzero values of entanglement and quantum coherence regarding the subspa\-ce of fermions, with $|\tilde{+}\rangle := \frac{1}{\sqrt{2}}\, ({f_1^{\dagger}} + {f_2^{\dagger}})\ket{\vac}$. In Figs.~\ref{fig:fig00007} and~\ref{fig:fig00008}, we show the concurrence and the $\ell_1$ norm of coherence in both the systems MFs$+$QD and RFs$+$QD, for the cutoff frequencies ${\omega_c} = 10$ and ${\omega_c} = 50$, respectively.

In Fig.~\ref{fig:fig00007}, for the case of RFs at zero temperature, we first note that the dynamics of entanglement and quantum co\-he\-ren\-ce are identical for each of the two-body reduced states of the system [see Figs.~\ref{fig:fig00007}(b),~\ref{fig:fig00007}(f), and~\ref{fig:fig00007}(j)]. In opposite, for MFs at zero temperature, concurrence of state ${\rho_{12}}(t)$ exhibits a revival after going to zero, and then decreases until completely va\-ni\-shing, while the quantum coherence oscillates until it reaches a stationary value [see Fig.~\ref{fig:fig00007}(a)]. In addition, Figs.~\ref{fig:fig00007}(e) and~\ref{fig:fig00007}(i) show the marginal states ${\rho_{1d}}(t)$ and ${\rho_{2d}}(t)$ have nonzero oscilla\-ting values of quantum coherence that approach a stationary value for a wide time window. In turn, concurrence of the state ${\rho_{1d}}(t)$ increases and suddenly goes to zero for short times [see Fig.~\ref{fig:fig00007}(e)], while for the state ${\rho_{2d}}(t)$ the concurrence shows revivals after going suddenly to zero, and then oscillates and remains finite for longer times [see Fig.~\ref{fig:fig00007}(i)]. Next, moving to the case of finite temperature ($\beta = 1$), the concurrence of state ${\rho_{12}}(t)$ decreases and goes to zero in both systems of MFs and RFs, also showing an intermediate revival before completely vanishing in the former setup, while quantum coherence starts decreasing and asymptotically converges to a stationary value [see Figs.~\ref{fig:fig00007}(c) and~\ref{fig:fig00007}(d)]. Furthermore, Figs.~\ref{fig:fig00007}(g),~\ref{fig:fig00007}(h),~\ref{fig:fig00007}(k), and~\ref{fig:fig00007}(l) show that states ${\rho_{1d}}(t)$ and ${\rho_{2d}}(t)$ have zero valued concurrence, except for a narrow peak that appears at short times of the dynamics, while the $\ell_1$ norm of coherence shows damped oscillations and then saturates to a fixed value.

In Fig.~\ref{fig:fig00008}, we show the dynamics in the Markovian regime with $\omega_c = 50$. We see that the curves of entanglement and quantum coherence are identical for the two-body reduced states of the setup comprising RFs at zero temperature [see Figs.~\ref{fig:fig00008}(b),~\ref{fig:fig00008}(f), and~\ref{fig:fig00008}(j)]. Figures~\ref{fig:fig00008}(f) and~\ref{fig:fig00008}(j) show the quantum resources for marginal states ${\rho_{1d}}(t)$ and ${\rho_{2d}}(t)$ exhibit rapid oscillations that are latter suppressed. For MFs at zero temperature, concurrence and quantum coherence of state ${\rho_{12}}(t)$ decays and saturates into a stationary value [see Fig.~\ref{fig:fig00008}(a)]. Furthermore, Figs.~\ref{fig:fig00008}(e) and~\ref{fig:fig00008}(i) show that the marginal states ${\rho_{1d}}(t)$ and ${\rho_{2d}}(t)$ have nonzero damped oscillating va\-lues of concurrence and quantum coherence, with the former going to zero and the later approaching a fixed nonzero value. For finite temperature ($\beta = 1$), Figs.~\ref{fig:fig00008}(c) and~\ref{fig:fig00008}(d) show the concurrence of state ${\rho_{12}}(t)$ decreases exponentially in the system of MFs, while abruptly going to zero in the case of RFs. In turn, quantum coherence of MFs starts decreasing and asymptotically converges to a stationary value, and the quantum coherence of RFs exhibits a revival after suddenly going to zero and then asymptotically approaches a stationary value. Figures~\ref{fig:fig00008}(g),~\ref{fig:fig00008}(h),~\ref{fig:fig00008}(k) and~\ref{fig:fig00008}(l) show the states ${\rho_{1d}}(t)$ and ${\rho_{2d}}(t)$ have nonzero quantum coherences with oscillation damping in the amplitudes, and also nonzero entanglement signaled by a narrow peak of concurrence that survives only for a short time window of the dynamics.

\paragraph{Werner initial state ---} We now consider the time evolution of the dynamics of a third type of single-fermion initial state. We choose ${{\rho}_S}(0) = |{W}\rangle\langle{W}|$, where $|{W}\rangle := \frac{1}{\sqrt{3}}\left({d^{\dagger}} + {f_1^{\dagger}} + {f_2^{\dagger}} \right)\ket{\vac}$ is the Werner state that fully correlates the two fermions and the QD, also exhibiting nonzero quantum coherence in all Hilbert subspaces. Figures~\ref{fig:fig00009} and~\ref{fig:fig00010} show the plots of concurrence and the $\ell_1$ norm of coherence for both the systems of MFs$+$QD and RFs$+$QD, for the cutoff frequencies ${\omega_c} = 10$ and ${\omega_c} = 50$, respectively. 

In Fig.~\ref{fig:fig00009}, for the case of RFs at zero temperature, values of concurrence and $\ell_1$ norm of coherence coincide for each of the two-body reduced states of the system [see Figs.~\ref{fig:fig00009}(b),~\ref{fig:fig00009}(f), and~\ref{fig:fig00009}(j)]. Conversely, for MFs at zero temperature, concurrence of states ${\rho_{jd}}(t)$ exhibits revivals after going to zero, while it reaches nonzero asymptotic values for the two-body states that mix both the QD and fermionic degrees of freedom [see Figs.~\ref{fig:fig00009}(e) and~\ref{fig:fig00009}(i)] and also vanishes at later times for the reduced state of two fermions [see Fig.~\ref{fig:fig00009}(a)]. In addition, for MFs at finite temperature, concurrence of the state ${\rho_{12}}(t)$ exhi\-bits revivals and asymptotically goes to zero [see Fig.~\ref{fig:fig00009}(c)], while abruptly vanishing for states ${\rho_{jd}}(t)$ [see Figs.~\ref{fig:fig00009}(g) and~\ref{fig:fig00009}(k)]. In turn, quantum coherence shows fluctuations in its amplitudes and approaches stationary values, regardless of the temperature [see Figs.~\ref{fig:fig00009}(a),~\ref{fig:fig00009}(c),~\ref{fig:fig00009}(e),~\ref{fig:fig00009}(g),~\ref{fig:fig00009}(i), and~\ref{fig:fig00009}(k)]. For the system of RFs, $\ell_1$ norm of coherence exhibits damped oscillations and vanishes asymptotically for the states ${\rho_{jd}}(t)$ [see Fig.~\ref{fig:fig00009}(h) and~\ref{fig:fig00009}(l)], while for the state ${\rho_{12}}(t)$ it reaches nonzero stationary value [see Fig.~\ref{fig:fig00009}(d)].

In the Markovian regime (${\omega_c} = 50$), at zero temperature ($T = 0$), Figs.~\ref{fig:fig00010}(b),~\ref{fig:fig00010}(f), and~\ref{fig:fig00010}(j) show the concurrence and quantum coherence of all states behaving identically in the system of RFs, experiencing rapid oscillations at earlier times. In contrast, for the system of MFs, both quantities take different values and exhibit nonmonotonic decays with rapid oscillations that are suppressed at later times [see Figs.~\ref{fig:fig00010}(a),~\ref{fig:fig00010}(e), and~\ref{fig:fig00010}(i)]. Next, moving to the case of finite temperature ($\beta = 1$), the concurrence of state ${\rho_{jd}}(t)$ decreases and suddenly goes to zero in both systems of MFs and RFs, while quantum coherence shows fluctuations in its amplitudes and asymptotically converges to a stationary value [see Figs.~\ref{fig:fig00010}(g),~\ref{fig:fig00010}(h),~\ref{fig:fig00010}(k), and~\ref{fig:fig00010}(l)]. In addition, for the marginal state ${\rho_{12}}(t)$, the quantum resources decay and approach stationary values [Figs.~\ref{fig:fig00010}(c) and~\ref{fig:fig00010}(d)]. We point out that quantum coherence converges to a finite value in the system MFs, while it goes to zero for the reduced state of two fermions. This shows that quantum co\-he\-ren\-ce of the two-body state of MFs is a more robust quantum resource than that of RFs.

\subsubsection{Two-fermions initial state}

We now present the dynamics of correlation and quantum coherences when the system is initialized in the initial state ${{\rho}_S}(0) = |\tilde{\phi}\rangle\langle\tilde{\phi}|$ that entangles two fermions, where $|\tilde{\phi}\rangle := \frac{1}{\sqrt{2}}(\mathbb{I} + {f_1^{\dagger}}{f_2^{\dagger}})\ket{\vac}$. Noteworthy, this initial state has even parity respective to the occupation number of fermions, also exhi\-bi\-ting nonzero correlations and quantum coherences between the two fermionic subspaces. In Figs.~\ref{fig:fig00011} and~\ref{fig:fig00012}, we compare the concurrence and the $\ell_1$ norm of coherence of the systems of MFs$+$QD and RFs$+$QD, for the cutoff frequencies ${\omega_c} = 10$ and ${\omega_c} = 50$, respectively. 

We shall begin discussing Fig.~\ref{fig:fig00011} for the case of non-Markovian dynamics (${\omega_c} = 10$). For MFs at zero temperature, Fig.~\ref{fig:fig00011}(a) shows that $\ell_1$ norm of coherence of ${\rho_{12}}(t)$ decreases, reaching a finite stationary value, while concurrence oscillates and suddenly goes to zero. In addition, Figs.~\ref{fig:fig00011}(e) and~\ref{fig:fig00011}(i) show the oscillation patterns for quantum coherence of states ${\rho_{1d}}(t)$ and ${\rho_{2d}}(t)$, with the latter saturating faster to a finite stationary value than the former. Interestingly, concurrence of state ${\rho_{1d}}(t)$ goes to zero after displaying a few revivals [see Fig.~\ref{fig:fig00011}(e)], while for the state ${\rho_{2d}}(t)$ concurrence starts increasing and exhibits periodic oscillations around a stationary value for larger times [see Fig.~\ref{fig:fig00011}(i)]. For RFs at zero temperature, Figs.~\ref{fig:fig00011}(f) and~\ref{fig:fig00011}(j) show that both quantum resources for states ${\rho_{jd}}(t)$ oscillate and vanish for larger times. In particular, note that concurrence and quantum coherence of state ${\rho_{1d}}(t)$ oscillates appro\-xi\-ma\-te\-ly with the same period [see Fig.~\ref{fig:fig00011}(f)], while the former goes to zero faster than the latter for the reduced state ${\rho_{2d}}(t)$ [see Fig.~\ref{fig:fig00011}(j)]. For the two-body state ${\rho_{12}}(t)$, Fig.~\ref{fig:fig00011}(b) shows that concurrence exhibits a revival after suddenly going to zero, while quantum coherence decays to a stationary value. Interestingly, both resources asym\-pto\-ti\-cally reach the same numerical value for later times of the dynamics. For MFs at finite temperature, Figs.~\ref{fig:fig00011}(g) and~\ref{fig:fig00011}(k) show the $\ell_1$ norm of coherence of states ${\rho_{jd}}(t)$ starts in\-crea\-sing and slowly reaches a finite stationary value, while concurrence shows narrow peaks at the earlier times of the dynamics. In Fig.~\ref{fig:fig00011}(c), concurrence of state ${\rho_{12}}(t)$ goes abruptly to zero, while $\ell_1$ norm of coherence decreases slowly. For RFs at finite temperature, Fig.~\ref{fig:fig00011}(d) shows the quantum coherence of state ${\rho_{12}}(t)$ decreases and remains finite, while concurrence suddenly goes to zero. In Fig.~\ref{fig:fig00011}(h), the concurrence of state ${\rho_{1d}}(t)$ shows a narrow peak at earlier times and suddenly goes to zero, while in Fig.~\ref{fig:fig00011}(l) concurrence of ${\rho_{2d}}(t)$ is zero at all times of the dynamics. For both two-body reduced states, the $\ell_1$ norm of coherence shows damped oscillations and approaches zero.

Next, let us comment on Fig.~\ref{fig:fig00012} for the case of Markovian dynamics (${\omega_c} = 50$). For MFs at zero temperature, Fig.~\ref{fig:fig00012}(a) shows that $\ell_1$ norm of coherence of ${\rho_{12}}(t)$ decreases exponentially, while concurrence suddenly goes to zero. In addition, Figs.~\ref{fig:fig00012}(e) and~\ref{fig:fig00012}(i) show the concurrence of states ${\rho_{1d}}(t)$ and ${\rho_{2d}}(t)$, with the former going to zero faster than the latter. In both cases, quantum coherence oscillates and approaches a stationary value for larger times. For RFs at zero temperature, Figs.~\ref{fig:fig00012}(f) and~\ref{fig:fig00012}(j) show that both quantum resources for states ${\rho_{jd}}(t)$ display rapid oscillations that are suppressed for later times. In contrast, for the two-body state ${\rho_{12}}(t)$, Fig.~\ref{fig:fig00012}(b) shows that concurrence exhibits a revival after dropping suddenly to zero, while quantum coherence decays  monotonically to a stationary value. We point out that both resources start saturating to the same numerical value for later times of the dynamics. For MFs at finite temperature, Figs.~\ref{fig:fig00012}(g) and~\ref{fig:fig00012}(k) show the states ${\rho_{jd}}(t)$ have zero valued concurrence, except for a narrow peak at the earlier times of the dynamics, while the $\ell_1$ norm of coherence shows damped oscillations and then saturates to a stationary value. In Fig.~\ref{fig:fig00012}(c), concurrence of state ${\rho_{12}}(t)$ abruptly goes to zero, while $\ell_1$ norm of coherence of state ${\rho_{12}}(t)$ monotonically decreases and va\-ni\-shes for larger times. For RFs at finite temperature, Fig.~\ref{fig:fig00012}(d) shows the quantum coherence of state ${\rho_{12}}(t)$  decreases monotonically and remains finite, while concurrence suddenly goes to zero. For the two-body reduced states ${\rho_{jd}}(t)$, Figs.~\ref{fig:fig00012}(h) and~\ref{fig:fig00012}(l) show the $\ell_1$ norm of coherence displays oscillations that are rapidly damped, while concurrence is zero at all times of the dynamics.

Before closing this section, a final remark is in order. Apart from the asymptotic case shown in Figs.~\ref{fig:fig00011}(b) and~\ref{fig:fig00012}(b), note that concurrence and quantum coherence display different dynamical behaviors, i.e., the two quantum resources do not coincide at any time of the dynamics. This feature somehow suggests a parity fingerprint that is related to the global occupation of the initial state. On the one hand, for initial states with odd parity in the occupation number, quantum coherence and concurrence coincide for some of the marginal states of the system (see Figs.~\ref{fig:fig00005},~\ref{fig:fig00006}, ~\ref{fig:fig00007},~\ref{fig:fig00008},~\ref{fig:fig00009} and~\ref{fig:fig00010}). On the other hand, Figs.~\ref{fig:fig00011} and~\ref{fig:fig00012} show the quantum resources behave quite differently at all times of the dynamics when the system is initialized in a probe state with even parity in the occupation number.

\begin{figure}[h!]
\begin{center}
\includegraphics[scale=0.925]{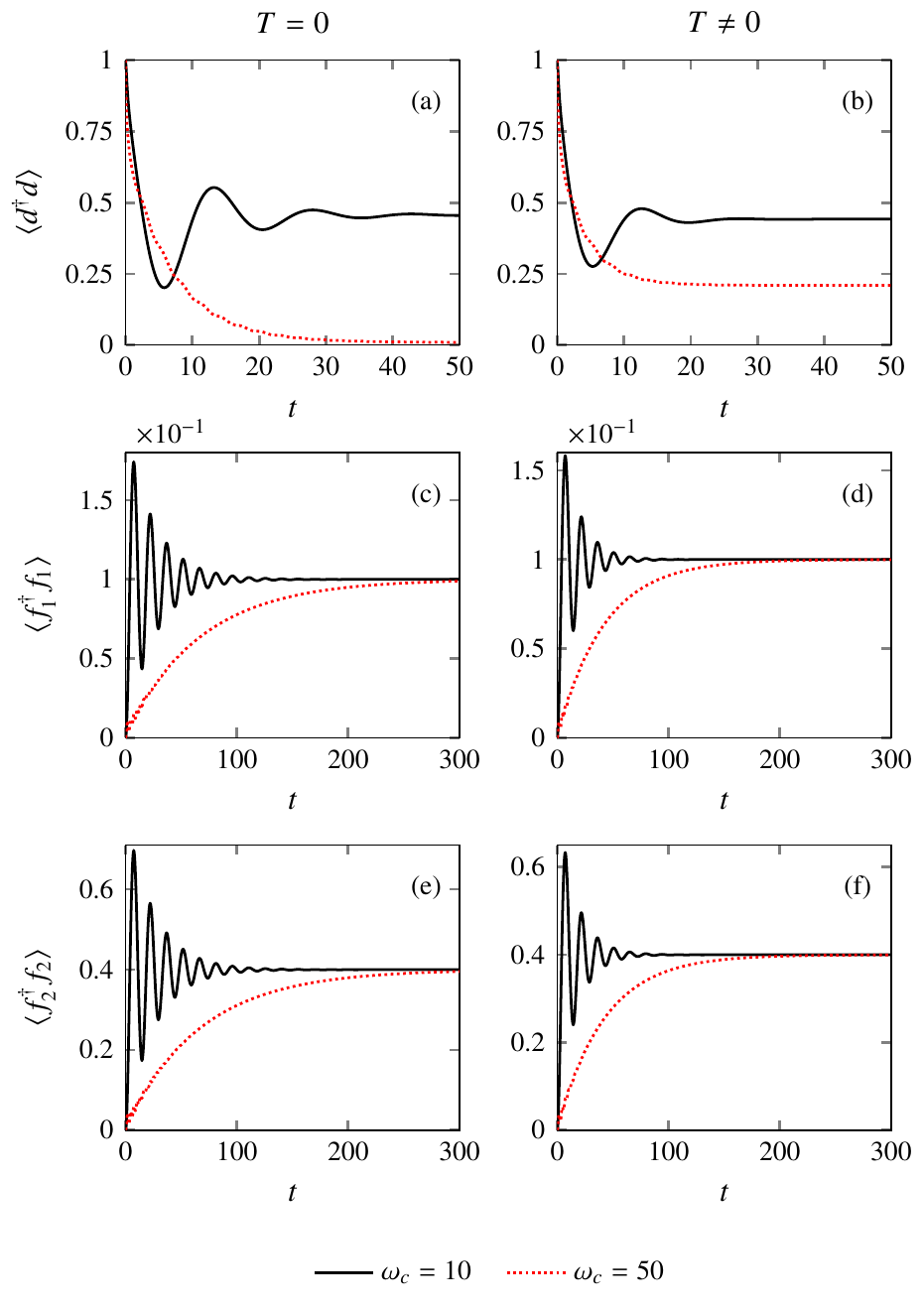}
\caption{(Color online) Populations of the subsystem of MFs$+$QD with nonlocal Majorana fermions ($\epsilon = 0.0005$), where $\langle \bullet \rangle = \text{Tr}(\bullet \, {\rho_S}(t))$, for both the non-Markovian (${\omega_c} = 10$) and Markovian dynamics (${\omega_c} = 50$), at zero ($T = 0$, right panels) and finite ($\beta = 1$, left panels) temperatures. The system of MFs$+$QD is initialized at the state ${{\rho}_S}(0) = |\tilde{1}\rangle\langle\tilde{1}|$, with $|\tilde{1}\rangle = {d^{\dagger}}\ket{\vac}$. Here we set $\gamma = 0.05$, $s = 1$, ${{\epsilon}_d} = 0.5$, and ${\lambda_2} = 2{\lambda_1} = 0.2$.}
\label{fig:fig00013}
\end{center}
\end{figure}

\begin{figure*}[t!]
\begin{center}
\includegraphics[scale=0.95]{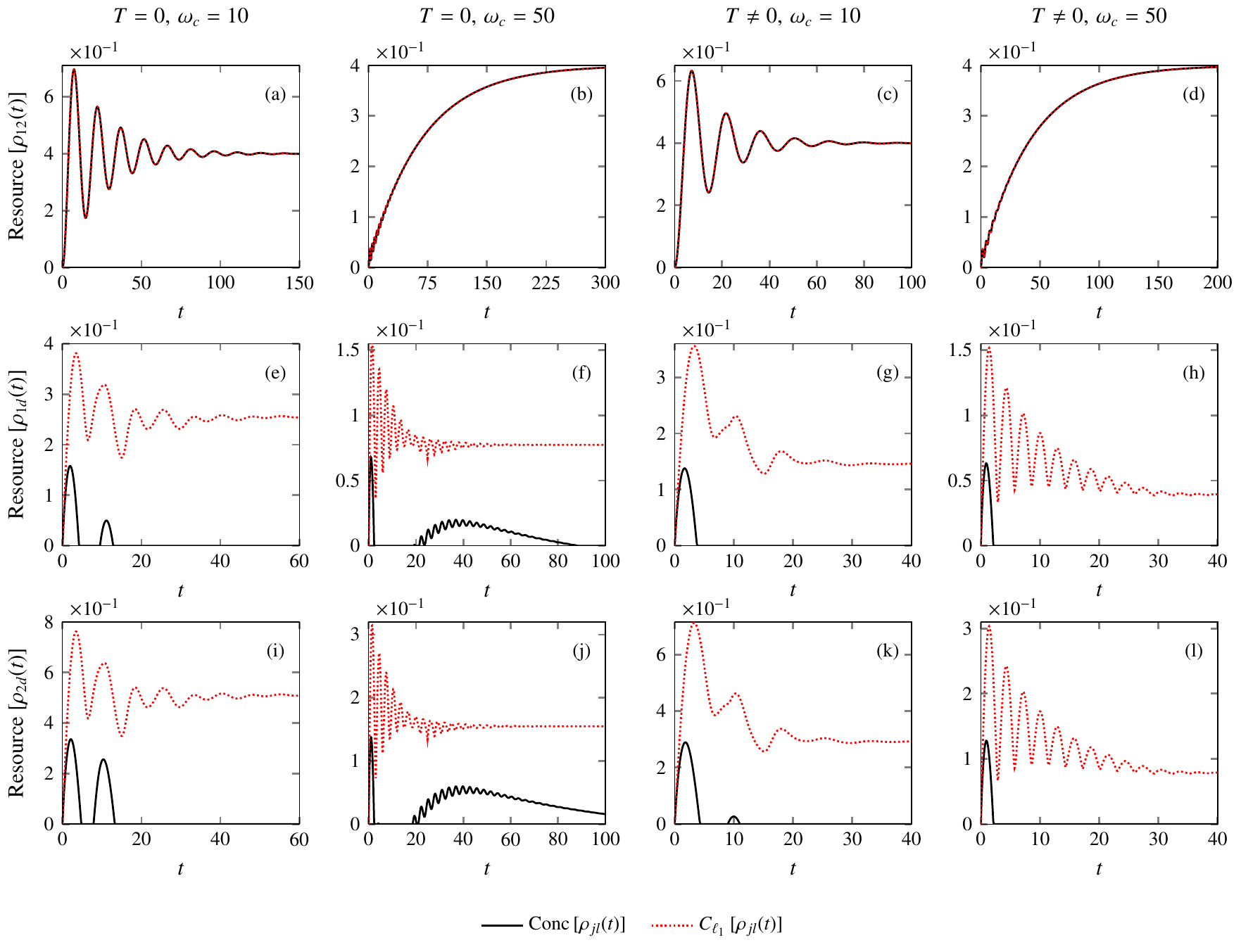}
\caption{(Color online) Plot of concurrence (black solid line) and $\ell_1$ norm of coherence (red dashed line) of the subsystem of MFs$+$QD with nonlocal Majorana fermions ($\epsilon = 0.0005$), for both the non-Markovian (${\omega_c} = 10$) and Markovian dynamics (${\omega_c} = 50$), at zero ($T = 0$) and finite ($\beta = 1$) temperatures. The system is initialized in the uncorrelated state ${{\rho}_S}(0) = |\tilde{1}\rangle\langle\tilde{1}|$, with $|\tilde{1}\rangle = {d^{\dagger}}\ket{\vac}$, with $\gamma = 0.05$, $s = 1$, ${{\epsilon}_d} = 0.5$, and ${\lambda_2} = 2{\lambda_1} = 0.2$.}
\label{fig:fig00014}
\end{center}
\end{figure*}


\section{Dynamics in the nonlocal regime of the Majorana fermions}
\label{sec:sec0005}

In this section we will discuss the possible role of nonlocality of Majorana fermions in the dynamics of occupations and quantum resources. The physical system of MFs$+$QD comprises a pair of Majorana bound states that arises at the ends of two superconductor nanowires of length $L$ [see Fig.~\ref{fig:fig00001}(a)]. In turn, for large values of $L$, i.e., taking the MFs far from apart each other, it is well known the couplings $\epsilon_j \approx 0$ ($j = 1,2$) between them will become negligible, mainly because such energies decrease exponentially with the length of the nanowires. Hence, in our setting, the smaller the coupling $\epsilon$, the more nonlocal should be the pair of MFs. In the following, we will investigate the nonlocality of MFs by setting $\epsilon = 0.0005$. For simplicity, the system of MFs$+$QD is initialized in the fully uncorrelated initial state ${{\rho}_S}(0) = |\tilde{1}\rangle\langle\tilde{1}|$, with $|\tilde{1}\rangle = {d^{\dagger}}\ket{\vac}$. Moreover, we set $\gamma = 0.05$, $s = 1$, ${{\epsilon}_d} = 0.5$, and ${\lambda_2} = 2{\lambda_1} = 0.2$.

Figure~\ref{fig:fig00013} shows the dynamics of the occupations in the system of MFs$+$QD. In the non-Markovian regime ($\omega_c = 10$), note the occupation of the QD and both the MFs take nonzero values and oscillate around stationary values, regardless of the temperature. For the case of Markovian dynamics ($\omega_c = 50$), we point out the occupation $\langle{d^{\dagger}}{d}\rangle$ mostly decreases, while $\{ \langle{f_j^{\dagger}}{f_j}\rangle\}_{j = 1,2}$ starts growing and approaches stationary values at later times of the dynamics, also regardless of the temperature of the system. In particular, note the occupation of the QD goes to zero at zero temperature [see Fig.~\ref{fig:fig00013}(a)], while it remains a nonzero constant for finite temperature ($\beta = 1$) [see Fig.~\ref{fig:fig00013}(b)].

In Fig.~\ref{fig:fig00014}, we plot the concurrence and the $\ell_1$ norm of coherence for the system of MFs$+$QD for the cutoff frequencies ${\omega_c} = 10$ and ${\omega_c} = 50$, at zero ($T = 0$) and finite temperature ($\beta = 1$). Note that entanglement and quantum coherence of the reduced state of fermions ${\rho_{12}}(t)$ take nonzero values, and also coincide for all times of the dynamics, regardless of temperature and cutoff frequencies [see Figs.~\ref{fig:fig00014}(a),~\ref{fig:fig00014}(b),~\ref{fig:fig00014}(c), and~\ref{fig:fig00014}(d)]. We point out the marginal states $\{ {\rho_{jd}}(t) \}_{j = 1,2}$ have nonzero oscillating quantum coherences that approach stationary values at later times, regardless the temperature, for both the non-Markovian ($\omega_c = 10$) and Markovian ($\omega_c = 50$) settings. However, the concurrence of such states is a zero valued quantity, except for narrow peaks that appear at earlier times of the dynamics as shown in Figs.~\ref{fig:fig00014}(e),~\ref{fig:fig00014}(g),~\ref{fig:fig00014}(i),~\ref{fig:fig00014}(k) ($T = 0$, $\omega_c = 10$), and Figs.~\ref{fig:fig00014}(h),~\ref{fig:fig00014}(l) ($\beta = 1$, $\omega_c = 50$). In particular, for $\beta = 1$ and $\omega_c = 10$, note that he concurrence of ${\rho_{1d}}(t)$ and ${\rho_{2d}}(t)$ shows an oscillation pattern that is smoothly suppressed to zero [see Figs.~\ref{fig:fig00014}(f) and~\ref{fig:fig00014}(j)].

Finally, we comment on the role of nonlocality of MFs into the quantum resources in comparison with the dynamics of such quantities for the system of local MFs (see Figs.~\ref{fig:fig00005} and~\ref{fig:fig00006}). In both cases, we observe qualitatively similar behavior of quantum coherence and entanglement, regardless of the dynamics and temperature effects. However, for the system composed of nonlocal MFs, the reduced states $\{ {\rho_{jd}}(t)\}_{j = 1,2}$ present more fragile entanglement to dissipative effects than the one that we found for local MFs. On the one hand, for the system of MFs$+$QD with nonlocal fermions undergoing the non-Markovian dynamics ($\omega_c = 10$) at zero temperature ($T = 0$), Figs.~\ref{fig:fig00014}(e) and~\ref{fig:fig00014}(i) show that the concurrence of the marginal states ${\rho_{1d}}(t)$ and ${\rho_{2d}}(t)$ exhibits only two nonzero peaks at earlier times of the dynamics. On the other hand, Figs.~\ref{fig:fig00005}(e) and~\ref{fig:fig00005}(i) show that, for the system of MFs$+$QD with local fermions at zero temperature, the entanglement stored in each reduced state $\{ {\rho_{jd}}(t)\}_{j = 1,2}$ asymptotically converges to a stationary value at later times of the non-Markovian dynamics ($\omega_c = 10$).


\section{Conclusions}
\label{sec:conclusions}

In this work we have investigated the dynamics of quantum resources in a tripartite fermionic system coupled to an external reservoir. Two classes of systems were considered: (i) a two-level quantum dot coupled to two regular fermion levels (RFs$+$QD) and (ii) a quantum dot coupled to two pairs of Majorana fermions (MFs$+$QD). In both cases, the quantum dot is coupled to a fermionic reservoir. Invoking a quantum master-equation approach that includes a memory kernel allowing study of both the Markovian to the non-Markovian regimes, at zero and finite temperatures, we analyze the time evolution of the quantum resources of the systems, namely, pairwise entanglement and quantum coherence, quantified by concurrence and ${{\ell}_1}$ norm, respectively. 

In general, we observe a clear distinction of the dynamics in each system, depending on the initial state and the pa\-ri\-ty of the global occupation of such state. On the one hand, for a fully uncorrelated incoherent initial state with a single fermion in the system, we observe that entanglement and coherence are generated by the dynamics at both the zero and finite temperatures in the MFs$+$QD system. On the other hand, in the RFs$+$QD these quantities are generated but survive only during a certain time window, vanishing at later times of the dynamics at zero temperature, and saturating at a small value at finite temperature. Overall, these features are observed in both the Markovian and non-Markovian dissipation regimes. 

For entangled and coherent single-fermion initial states, the quantum resources decrease but saturate to finite stationary values for the MFs$+$QD system regardless of temperature, for both the Markovian and non-Markovian regimes. On the other hand, for the RFs$+$QD system, the quantum coherence settles down at a finite value at nonzero temperature. Similar features are observed for choosing an initial state with two fermions in the system. We discussed the physical setting in which the Majorana fermions in each pair are almost fully decoupled from each other, i.e., typically a long topological nanowire. We observe qualitatively similar behavior, showing the robustness of our results against changes in the to\-po\-lo\-gical superconductor hosting the Majorana fermions. 

Finally, we highlight the correlated behavior of the quantum resources at both zero and finite temperatures in the MFs$+$QD system, which is not observed in the RFs$+$QD case. This might be seen as a signature of the property called genuine and distributed correlated coherence, which in turn witnesses the amount of quantum coherence that is contained within the correlations of the tripartite system~\cite{PhysRevA.94.022329,JMathPhys_51_414013}. As a final remark, we point out that the Hamiltonian in Eq.~\eqref{eq:eq000006} can be extended to a more realistic scenario that includes the coupling of MFs on different wires. To do so, Eq.~\eqref{eq:eq000006} should be recast to include the contribution of parity-dependent splitting terms to take in account the imbalance of phases related to tunneling matrix elements (see, for example, Refs.~\cite{PhysRevLett.106.090503,PhysRevB.95.235305}). For this case, one could investigate the robustness of quantum coherence and entanglement to temperature fluctuations and environmental decoherence. Indeed, this is an issue that we hope to address in further investigations.


\begin{acknowledgments}
We thank I. de Vega for fruitful conversations. The authors acknowledge the financial support from the Brazilian ministries MEC and MCTIC. The project was funded by Brazilian funding agencies CNPq (Grant No. 307028/2019-4 and Grant No. 305738/2018-6), FAPESP (Grant No. 2017/03727-0), Coordena\c{c}\~{a}o de Aperfei\c{c}oamento de Pessoal de N\'{i}vel Superior -- Brasil (CAPES) (Finance Code 001), and by the Brazilian National Institute of Science and Technology for Quantum Information (INCT-IQ) Grant No. 465469/2014-0.
\end{acknowledgments}

\setcounter{equation}{0}
\setcounter{figure}{0}
\setcounter{table}{0}
\setcounter{section}{0}
\numberwithin{equation}{section}
\makeatletter
\renewcommand{\thesection}{\Alph{section}} 
\renewcommand{\thesubsection}{\thesection.\arabic{subsection}}
\renewcommand{\theequation}{\Alph{section}\arabic{equation}}
\renewcommand{\thefigure}{\arabic{figure}}
\renewcommand{\bibnumfmt}[1]{[#1]}
\renewcommand{\citenumfont}[1]{#1}


\section*{Appendix}


\section{Details on the Master equation}
\label{sec:appendix0A}

The reduced dynamics of the system MFs$+$QD is described by the quantum master equation
\begin{align}
\label{eq:eq000024}
\frac{d{\rho_S}(t)}{dt} &= - i\, [{H_S},{\rho_S} (t) ] + {\int_0^t} d\tau  {{\alpha}^+}(t - \tau) [({V_{\tau - t}} \, {d^{\dagger}}) {\rho_S} (t) ,d] \nonumber\\
&+ {\int_0^t} d\tau {{\alpha}^-}(t - \tau)[({V_{\tau - t}} \, d) \, {\rho_S} (t),{d^{\dagger}}] + \text{H.c.} ~,
\end{align}
with ${H_S} = {H_1} + {H_2} + {H_3}$, where ${H_1} = {\epsilon_d}{\hat{n}_d}$ is the QD Hamiltonian, ${H_2} = {\sum_{j = 1}^2}\, {\epsilon_j}\left({\hat{n}_j} - {1}/{2}\right)$ is the MF (RF) Hamiltonian, and ${H_3} = {\sum_{j = 1,2}}({\lambda_j}\,{d^{\dagger}}{f_j} + {\widetilde{\lambda}_j}\,{d^{\dagger}}{f_j^{\dagger}} + {\rm H.c.})$ stands for the Hamiltonian modeling the MFs$+$QD (RFs$+$QD) coupling. To clarify, by setting ${\widetilde{\lambda}_j} = {\lambda_j}$ one obtains the Hamiltonian of the system MFs$+$QD, while the Hamiltonian of the system RFs$+$QD is recovered imposing that ${\widetilde{\lambda}_j} = 0$. In the former case, the e\-ner\-gies $\{ {E_j} \}_{j = 1,\ldots,8}$ and the eigenstates $\{ | {E_j}\rangle\}_{j = 1,\ldots, 8}$ of the system MFs$+$QD have been presented in Table~\ref{tab:tab000001}, while for the latter the spectral decomposition of the Hamiltonian for RFs$+$QD is given in Table~\ref{tab:tab000002} in the main text. In Eq.~\eqref{eq:eq000024} we have introduced the operator ${V_{\tau - t}}\, {d^{\dagger}} = {e^{ i (\tau - t) {H_S}}} {d^{\dagger}} \, {e^{- i (\tau - t) {H_S}}}$, which in turn can be written more conveniently as
\begin{align}
\label{eq:eq000025}
{{V}_{\tau - t}}\,{d^{\dagger}} = {\sum_{j,l = 1}^8} \, {{e}^{i(\tau - t)({{E}_j} - {{E}_l})}} \,  \langle{E_j}|{{d}^{\dagger}}|{E_l}\rangle \, |{E_j}\rangle\langle{E_l}| ~,
\end{align}
with
\begin{equation}
\label{eq:eq000026}
 \langle{E_j}|{{d}^{\dagger}}|{E_l}\rangle = {\sum_{{k_1}, \, {k_2}}}  \, {{(-1)}^{ {{k}_1} + {{k}_2} }} \langle{E_j}|{{k}_1},{{k}_2}, 1\rangle\langle{{k}_1},{{k}_2},0|{E_l}\rangle ~.
\end{equation}
Next, the system-bath correlation functions are given by
\begin{equation}
\label{eq:eq000027}
{\alpha^+} (t) = {\int_0^{\infty}} d\omega \, J(\omega) {N_F}(\omega) \, {e^{i\omega t}} ~, 
\end{equation}
and
\begin{equation}
\label{eq:eq000028}
{\alpha^-} (t) = {\int_0^{\infty}} d\omega \, J(\omega)({N_F}(\omega) + 1)\, {e^{-i\omega t}} ~,
\end{equation}
with ${N_F}({\omega}) = {{[\exp(\beta{\omega})+1]}^{-1}}$ being the Fermi-Dirac distribution related to the fermionic reservoir and $J(\omega) = \gamma \, {{\omega}^s} \, {{\omega}_c^{1 - s}} {e^{-\omega/{{\omega}_c}}}$ being the spectral density of the environment. For all $s > 0$ and finite temperature $0 < T < \infty$ (i.e., $\infty > \beta > 0$), it is straightforward to verify that
\begin{equation}
\label{eq:eq000029}
{{{{\alpha}^+}(t)}} =  \frac{\gamma}{4\, {\beta^2}} \, {(2\beta{\omega_c})^{1 - s}} \, \Gamma(1 + s)  \left[ {{\xi}_{1 + s}}\left( z(t)\right) - {{\xi}_{1 + s}}\left( z(t) + 1/2 \right) \right] ~,
\end{equation}
with 
\begin{equation}
\label{eq:eq000030}
z(t) := \frac{1 + \beta{{\omega}_c} - i{\omega_c}t}{2\beta{{\omega}_c}} ~,
\end{equation}
and
\begin{equation}
\label{eq:eq000031}
{{{{\alpha}^-}(t)}} =  {{{{\alpha}^+}(t)}^*} + \frac{\gamma\, {{\omega}_c^2}\, \Gamma(1 + s)}{ {\left( 1 + i{{\omega}_c} t \right)}^{1 + s} }   ~,
\end{equation}
where ${{\xi}_{1 + s}}(x)$ defines the generalized Riemann zeta function as ${{\xi}_{1 + s}}(x) = {\sum_{k = 0}^{\infty}} \, { {{(x + k)}^{- 1 - s}} } $. Noteworthy, for the asymptotic case of zero temperature $T \rightarrow 0$ (i.e., $\beta \rightarrow \infty$), one gets ${\lim_{\beta \rightarrow \infty}} \, {{{{\alpha}^+}(t)}} = {\lim_{\beta \rightarrow \infty}} \, {{{{\alpha}^+}(t)}^*} = 0$, and thus
\begin{equation}
\label{eq:eq000032}
{\lim_{\beta \rightarrow \infty}} \, {{\alpha^-}(t)} = \frac{\gamma\, {{\omega}_c^2}\, \Gamma(1 + s)}{ {\left( 1 + i{{\omega}_c}t \right) }^{1 + s} }   ~.
\end{equation}

In the following we will discuss how to solve the master equation in Eq.~\eqref{eq:eq000024} for the reduced density matrix ${{\rho}_S}(t)$. To do so, we point out the marginal state ${{\rho}_S}(t)$ can be written in terms of the occupation number basis states $\{|{n_1},{n_2},{n_d}\rangle\}$ as follows
\begin{equation}
\label{eq:eq000033}
{{\rho}_S}(t) = {\sum_{ \mathbf{k}, \mathbf{m} }} \, {{A}^{ {k_1}, \, {k_2}, \, {k_d}}_{ {m_1}, \, {m_2},\, {m_d} }}(t)\,  |{k_1},{k_2},{k_d}\rangle\langle{m_1},{m_2},{m_d}| ~,
\end{equation}
where $\mathbf{k} = ({k_1},{k_2},{k_d} )$, $\mathbf{n} = ( {m_1},{m_2},{m_d})$, with ${k_j} = \{0,1\}$, and ${m_j} = \{0,1\}$ for $j = \{1,2,d\}$, while
\begin{equation}
\label{eq:eq000034}
{{A}^{ {k_1},\,{k_2},\, {k_d}}_{ {m_1},\, {m_2},\, {m_d} }}(t) = \langle{k_1},{k_2},{k_d}|{{\rho}_S}(t)|{m_1},{m_2},{m_d}\rangle ~.
\end{equation}
Next, we will substitute Eq.~\eqref{eq:eq000033} into Eq.~\eqref{eq:eq000024}, and also project the aforementioned master equation onto the occupation number basis, thus obtaining a set of coupled differential equations for the time-dependent coefficients $\{ {{A}^{ {k_1}, \, {k_2}, \, {k_d}}_{ {m_1}, \, {m_2},\,{m_d} }}(t) \}_{\mathbf{k}, \mathbf{m}}$. By proceeding in this way, one gets
\begin{widetext}
\begin{align}
\label{eq:eq000035}
& \frac{d}{dt}{{A}^{ {\ell_1},\, {\ell_2},\,{\ell_d}}_{ {n_1},\,{n_2},\,{n_d} }}(t) = - i\, \langle{{\ell}_1},{{\ell}_2},{{\ell}_d}|[{ {{H}_S}, {{\rho}_S}(t)}]|{{n}_1},{{n}_2},{{n}_d}\rangle + {\int_0^t}\, d\tau \, {{\alpha}^+}(t - \tau) \, \langle{{\ell}_1},{{\ell}_2},{{\ell}_d}|[{ ({{V}_{\tau - t}}\,{{d}^{\dagger}}) \, {{\rho}_S}(t), d }]|{{n}_1},{{n}_2},{{n}_d}\rangle \nonumber\\ 
&+ {\int_0^t}\, d\tau \, {{\alpha}^-}(t - \tau) \, \langle{{\ell}_1},{{\ell}_2},{{\ell}_d}|[{ ({{V}_{\tau - t}}\,{d}) \, {{\rho}_S}(t), {{d}^{\dagger}} }]|{{n}_1},{{n}_2},{{n}_d}\rangle + \text{H.c.} ~.
\end{align}
To solve the set of differential equations in Eq.~\eqref{eq:eq000035}, we require numerical simulations. We see that, for a given initial state ${\rho_S}(0)$, Eqs.~\eqref{eq:eq000033} and~\eqref{eq:eq000035} fully characterize the marginal state ${\rho_S}(t)$. We point out that it is possible to simplify each contribution that appears in the right-hand side of Eq.~\eqref{eq:eq000035}. In this regard, the matrix elements of the sum of Hamiltonians $H_1 + H_2$ respective to the occupation number basis are given by
\begin{align}
\label{eq:eq000036}
\langle{{\ell}_1},{{\ell}_2},{{\ell}_d}|[{ {H_1} + {H_2}, {{\rho}_S}(t)}]|{{n}_1},{{n}_2},{{n}_d}\rangle = {\sum_{j = \{1,2,d\}}} {{\epsilon}_j} \, ({{\delta}_{{{\ell}_j},1}} {{\ell}_j} - {{\delta}_{{{n}_j},1}} {{n}_j}) \, {A^{{\ell_1},\,{\ell_2},\,{\ell_d}}_{{n_1},\, {n_2},\, {n_d}}}(t) ~,
\end{align}
while the matrix element for the coupling Hamiltonian $H_3$ reads as
\begin{align}
\label{eq:eq000037}
\langle{{\ell}_1},{{\ell}_2},{{\ell}_d}|[{ {{H}_3}, {{\rho}_S}(t)}]|{{n}_1},{{n}_2},{{n}_d}\rangle &= {\lambda_1} {\sum_{j = 0,1}}\, {(-1)^j} \left[  {{(-1)}^{{\ell_2}}} \, {\mathcal{C}_{{\ell_1},\, {\ell_d},j}} \, {A^{{\ell_1} - 2j + 1,\, {\ell_2},\, {\ell_d} + 2j - 1}_{{n_1},\,{n_2},\,{n_d}}} (t) - {{(-1)}^{ {n_2}}} \, {\mathcal{C}_{{n_1},\, {n_d}, j}} \, {A_{{n_1} - 2j + 1,\, {n_2},\, {n_d} + 2j - 1}^{{\ell_1},\,{\ell_2},\,{\ell_d}}} (t) \right] \nonumber\\
& + {\widetilde{\lambda}_1} {\sum_{j = 0,1}}\, {(-1)^j} \left[ {(-1)^{\ell_2}} \, {\widetilde{\mathcal{C}}_{{\ell_1},\, {\ell_d},j}} \, {A^{{\ell_1} + 2j - 1,\, {\ell_2},\, {\ell_d} + 2j - 1}_{{n_1},\,{n_2},\,{n_d}}} (t) -  {{(-1)}^{{n_2}}} \, {\widetilde{\mathcal{C}}_{{n_1},\, {n_d},j}} \, {A_{{n_1} + 2j - 1,\, {n_2},\, {n_d} + 2j - 1}^{{\ell_1},\,{\ell_2},\,{\ell_d}}} (t) \right] \nonumber\\
& + {\lambda_2} {\sum_{j = 0,1}}\, {(-1)^j} \left[ {\mathcal{C}_{{\ell_2},\, {\ell_d},j}} \, {A^{{\ell_1}, \, {\ell_2} - 2j + 1,\, {\ell_d} + 2j - 1}_{{n_1},\,{n_2},\,{n_d}}} (t) -  {\mathcal{C}_{{n_2},\, {n_d},j}} \, {A_{{n_1}, \, {n_2} - 2j + 1,\, {n_d} + 2j - 1}^{{\ell_1},\,{\ell_2},\,{\ell_d}}} (t) \right] \nonumber\\
& + {\widetilde{\lambda}_2} {\sum_{j = 0,1}}\, {(-1)^j} \left[ {\widetilde{\mathcal{C}}_{{\ell_2},\, {\ell_d},j}}  \, {A^{{\ell_1},\, {\ell_2} + 2j - 1,\, {\ell_d} + 2j - 1}_{{n_1},\,{n_2},\,{n_d}}} (t) - {\widetilde{\mathcal{C}}_{{n_2},\, {n_d},j}}  \, {A_{{n_1},\, {n_2} + 2j - 1,\, {n_d} + 2j - 1}^{{\ell_1},\,{\ell_2},\,{\ell_d}}} (t) \right] ~,
\end{align}
\end{widetext}
where we have defined the time-independent coefficients
\begin{equation}
\label{eq:eq000038}
{\mathcal{C}_{x,y,j}} = {\delta_{x,j}} \, {\delta_{y, 1 - j}} \, {(-1)^x} \sqrt{(x + 1 - j)(y + j)} ~,
\end{equation}
and
\begin{equation}
\label{eq:eq000039}
{\widetilde{\mathcal{C}}_{x,y,j}} = {\delta_{x,1 - j}} \, {\delta_{y, 1 - j}} \, {(-1)^x} \sqrt{(x + j)(y + j)} ~.
\end{equation}
We point out that Eq.~\eqref{eq:eq000037} will reduce to a simpler form for the case of Majorana fermions (${\widetilde{\lambda}_j} = {\lambda_j}$), and also for regular fermions (${\widetilde{\lambda}_j} = 0$). Moreover, one may verify that
\begin{align}
\label{eq:eq000040}
& {\int_0^t}\, d\tau \, {{\alpha}^+}(t - \tau) \, \langle{{\ell}_1},{{\ell}_2},{{\ell}_d}|[{ ({{V}_{\tau - t}}\,{{d}^{\dagger}}) \, {{\rho}_S}(t), d }]|{{n}_1},{{n}_2},{{n}_d}\rangle = \nonumber\\
=& {\sum_{j,l = 1}^8}\, {{\texttt{G}}_{jl}^{+}}(t) \, \langle{E_j}|\, {{d}^{\dagger}}|{E_l}\rangle \left[ {\delta_{{n_d} , 1}} {{(-1)}^{ {{n}_1} + {{n}_2} }} \sqrt{n_d} \, {\mathcal{Y}^{{\ell_1},\,{\ell_2},\,{\ell_d}}_{{n_1},\, {n_2},\, {n_d} - 1}}(j,l,t) \right. \nonumber\\
&\left. - \, {\delta_{{\ell_d} , 0}} \, {{(-1)}^{ {{\ell}_1} + {{\ell}_2} }} \sqrt{{{\ell}_d} + 1} \, {\mathcal{Y}^{{\ell_1},\,{\ell_2},\,{\ell_d} + 1}_{{n_1},\, {n_2},\, {n_d}}}(j,l,t) \right]~,
\end{align}
and
%
\begin{align}
\label{eq:eq000041}
& {\int_0^t}\, d\tau \, {{\alpha}^-}(t - \tau) \, \langle{{\ell}_1},{{\ell}_2},{{\ell}_d}|[{ ({{V}_{\tau - t}}\,{d}) \, {{\rho}_S}(t), {{d}^{\dagger}} }]|{{n}_1},{{n}_2},{{n}_d}\rangle = \nonumber\\
& = {\sum_{j,l = 1}^8}\, {{\texttt{G}}_{jl}^{-}}(t) \, \langle{E_j}|{d}|{E_l}\rangle \left[ {\delta_{{n_d} , 0}} \, {{(-1)}^{ {{n}_1} + {{n}_2} }} \sqrt{{{n}_d} + 1} \, {\mathcal{Y}^{{\ell_1},\,{\ell_2},\,{\ell_d}}_{{n_1},\, {n_2},\, {n_d} + 1}}(j,l,t) \right. \nonumber\\
&\left. - \, {\delta_{{\ell_d} , 1}} {{(-1)}^{ {{\ell}_1} + {{\ell}_2} }} \sqrt{{{\ell}_d}} \, {\mathcal{Y}^{{\ell_1},\,{\ell_2},\,{\ell_d} - 1}_{{n_1},\, {n_2},\, {n_d}}}(j,l,t) \right] ~,
\end{align}
where we define 
\begin{equation}
\label{eq:eq000042}
{{\texttt{G}}_{jl}^{\pm}}(t) := {\int_0^t}\, d\tau \, {{\alpha}^{\pm}}(t - \tau) \, {{e}^{i(\tau - t)({{E}_j} - {{E}_l})}} ~,
\end{equation}
and
\begin{equation}
\label{eq:eq000043}
{\mathcal{Y}^{{\ell_1},\,{\ell_2},\,{\ell_d}}_{{n_1},\, {n_2},\, {n_d}}}(j,l,t) := {\sum_{ {k_1},\, {k_2}, \, {k_d}}}  \langle{{\ell}_1},{{\ell}_2},{{\ell}_d}|{E_j}\rangle\langle{E_l}|{k_1},{k_2},{k_d}\rangle \, {A^{{k_1},\,{k_2},\,{k_d}}_{{n_1},\, {n_2},\, {n_d}}}(t) ~,
\end{equation}
with the matrix element $ \langle{E_j}|{d^{\dagger}}|{E_l}\rangle$ explicitly defined in Eq.~\eqref{eq:eq000026}.

\end{document}